\documentclass[useAMS,usenatbib]{mn2e}
\usepackage{graphicx}
\usepackage{multirow}
\def\f9{Fairall\,9}
\def\fa{Fe K$\alpha$~}
\def\fb{Fe K$\beta$~}
\def\aj{AJ}%
\def\araa{ARA\&A}%
\def\apj{ApJ}%
\def\apjl{ApJ}%
\def\apjs{ApJS}%
\def\aap{A\&A}%
\def\aaps{A\&AS}%
\def\mnras{MNRAS}%
\title[\textit{XMM-Newton} observations of \f9]{An \textit{XMM-Newton} view of the `bare' nucleus of \f9\thanks{Based on observations obtained with \textit{XMM-Newton}, an ESA science mission with instruments and contributions directly funded by ESA Member States and NASA.}}
\author[D.~Emmanoulopoulos et al.]{D.~Emmanoulopoulos,$^{1}$\thanks{E-mail: D.Emmanoulopoulos@soton.ac.uk} I.~E.~Papadakis,$^{2,3}$ I.~M.~M\textsuperscript{c}Hardy,$^{1}$ F.~Nicastro,$^{3,4}$ \newauthor S.~ Bianchi$^{3,5,6}$ and P.~Ar\'{e}valo$^{7}$ \\
$^{1}$School of Physics and Astronomy, University of Southampton, SO17 1BJ Southampton, United Kingdom\\
$^{2}$Physics Department, University of Crete, PO Box 2208, 71003 Heraklion, Greece\\
$^{3}$IESL, Foundation for Research and Technology, 71110 Heraklion, Greece\\
$^{4}$INAF-Osservatorio Astronomico di Roma, Via di Frascati 33, 00040 Monte Porzio Catone, Italy\\ 
$^{5}$Dipartimento di Fisica, Universit\`{a} degli Studi Roma Tre, via della Vasca Navale 84, 00146 Roma, Italy\\
$^{6}$INAF-Osservatorio Astronomico di Brera, Via E. Bianchi 46, 23807 Merate, Italy\\
$^{7}$Departamento de Ciencias Fisicas, Universidad Andres Bello, Av.Republica 252, Santiago, Chile}
\begin{document}

\date{Accepted 2011 March 31. Received 2011 March 30; in original form 2011 January 27}
\pagerange{\pageref{firstpage}--\pageref{lastpage}} \pubyear{2002}
\maketitle

\label{firstpage}
\begin{abstract}
We present the spectral results from a 130 ks observation, obtained from the \textit{X-ray Multi-Mirror Mission-Newton} (\textit{XMM-Newton}) observatory, of the type $\rmn{I}$ Seyfert galaxy \f9. An X-ray hardness-ratio analysis of the light-curves, reveals a `softer-when-brighter' behaviour which is typical for radio-quiet type $\rmn{I}$ Seyfert galaxies. Moreover, we analyse the high spectral-resolution data of the reflection grating spectrometer and we did not find any significant evidence supporting the presence of warm-absorber in the low X-ray energy part of the source's spectrum. This means that the central nucleus of \f9 is `clean' and thus its X-ray spectral properties probe directly the physical conditions of the central engine. The overall X-ray spectrum in the 0.5--10 keV energy-range, derived from the EPIC data, can be modelled by a relativistically blurred disc-reflection model. This spectral model yields for \f9 an intermediate black-hole best-fit spin parameter of $\alpha=0.39^{+0.48}_{-0.30}$.
\end{abstract}


\section{Introduction}
\label{sect:intro}
\f9 \citep{fairall77} is a type $\rmn{I}$ Seyfert galaxy at a redshift of z=0.047 \citep{lauberts89}. It has a black-hole (BH) mass of $(2.55\pm0.56)\times 10^8\;{\rm M}_{\sun}$ \citep[based on reverberation mapping,][]{peterson04} and an Eddington luminosity fraction $L_{\rm Bol}/L_{\rm Edd}$ of 0.05, assuming the bolometric luminosity, $L_{\rm Bol}=1.7\times10^{45}$ erg s$^{-1}$, after \citet{woo00}.\par
From the early 70's \f9 \citep[also known as ESO\,113-IG\,45,][]{holmberg78} has been observed several times in the X-ray band with \textit{Uhuru} \citep{forman78}, \textit{Ariel $\rmn{V}$} \citep{mchardy81}, \textit{HEAO-1\,A1} \citep{wood84}, \textit{HEAO-1\,A2} \citep{piccinotti82}, \textit{Einstein} \citep{petre84} and \textit{EXOSAT} \citep{morini86}. The \textit{Advanced Satellite for Cosmology and Astrophysics} (\textit{ASCA}) revealed a broad iron line which was well modelled by a relativistic line-emission model \citep{reynolds97}. Further observations with the \textit{X-ray Multi-Mirror Mission-Newton} (\textit{XMM-Newton}) \textit{observatory} \citep{gondoin01} unveiled the existence of an absorption edge at 7.64 keV (in the source's rest-frame). In contrast to the \textit{ASCA} observations, the iron line appeared to be narrow in the \textit{XMM-Newton} data-set, suggesting that the reflection process arises from material far from the putative central BH. \citet{brenneman09} reanalysed the same observation and found a third narrow emission-line at 6.78 keV (in the source's rest-frame) corresponding to ionized iron.\par
Additionally, recent \textit{Suzaku} observations of \f9 showed a broad asymmetric iron line \citep{schmoll09}, which is consistent with a disc-origin. A relativistically blurred disc reflection model, was fitted to the broad-band \textit{Suzaku} spectrum (0.7--30 keV) of \f9, yielding a constraint for the BH spin parameter of $\alpha=0.60\pm0.07$, where $\alpha=J M^{-2}$ \citep[in the geometric unit system,][]{bardeen72} and $-1<\alpha<1$, with $J$ being the angular momentum of the BH and $M$ the BH mass. Finally, \citet{patrick11} reanalysed the same data using additionally data from the burst alert telescope (BAT) on board \textit{SWIFT} satellite and derived a similar low value of $\alpha=0.44^{+0.04}_{-0.11}$.\par
In the current paradigm, active galactic nuclei (AGN) are thought to consist of an accretion disc surrounding a supermassive BH \citep{rees84}. Consequently, most of the emitted energy that is released in the X-rays is the result of the conversion of gravitational potential energy from matter falling towards the BH.\par
Thus, the X-ray spectroscopic properties of AGN can disclose valuable information about the physical properties of the processes operating in the immediate environment of the BH. In particular, broadening of the iron emission-lines, formed in the inner part of the accretion disc, can provide constraints about the BH spin parameter \citep{miller07}. Since the X-ray nucleus in \f9 is `clean', lacking X-ray warm-absorption features \citep{gondoin01}, it offers an ideal environment to investigate the physical properties of the central region.\par
However, there are alternative scenarios that can explain the broadening of the iron lines. Complex absorption processes from warm-absorbers can cause an apparently similar iron line profile to the profiles caused by strong gravity environments  \citep{reeves04,turner05}. Moreover, it is reasonable to expect several weak helium- and hydrogen-like emission-lines from distant material around the same energy as the iron line (i.e.\ 6.4--6.7 keV energy-range in the source's rest-frame), causing similar apparent broadening effects \citep{nandra07}. Also \citet{nandra07} showed that intermediate ionization species of the iron itself, can produce asymmetric broad line-profiles extending up to 6.97 keV (in the source's rest-frame). Finally, blueshifted or redshifted spectral-features from ejected or infalling materiel, close to the nucleus, can appear in the energies around the iron line, broadening its profile \citep{yaqoob99,turner02}.\par      
In this paper we report the analysis results of an \textit{XMM-Newton} observation of \f9 performed in December 2009. Initially, in Sect.~\ref{sect:obs_reduc} we present the data-reduction procedures for the X-ray data from the European photon imaging camera (EPIC), consisting of the pn-charge coupled device (pn-CCD) and the two metal oxide semi-conductor (MOS)-CCDs. In the same section, we present the data-reduction details for the soft X-ray data from the reflecting grating spectrometer (RGS). Then, in Sect.~\ref{sect:lcs} we show the EPIC light-curve products together with a hardness-ratio analysis. In Sect.~\ref{sect:rgs_spec}, we perform a spectral analysis of the RGS data and in Sect.~\ref{sect:epic_spec} we investigate the spectral properties of the EPIC data. A discussion together with a summary of our results can be found in Sect.~\ref{sect:discus}. The cosmological parameters used throughout this paper are: $\rm{H}_0=70$ km s$^{-1}$ Mpc$^{-1}$, $\Omega_{\Lambda}$=0.73 and $\Omega_{\rm m}$=0.27, yielding a luminosity distance to \f9 of 208.5 Mpc.

\section{OBSERVATIONS AND DATA-REDUCTION}
\label{sect:obs_reduc}
\subsection{EPIC-MOS and EPIC-pn data-reduction}
\label{ssect:xray_obs_reduc}
\f9 has been observed by \textit{XMM-Newton} (Obs-ID: 0605800401) from 2009 December 9, 19:56:35 (UTC), to 2009 December 11, 08:04:07 (UTC) (on-time: 130052 s). The pn camera was operated in Prime Small Window mode and the two MOS cameras in Prime Partial W2 mode. Medium-thickness aluminised optical blocking filters were used for all EPIC cameras to reduce the contamination of the X-rays from infrared, visible, and ultra-violet light.\par
The EPIC raw-data are reduced with the \textit{XMM-Newton} {\sc Scientific Analysis System} ({\sc SAS}) \citep{gabriel04} version 10.0.2. After reprocessing the pn and the two MOS data-sets with the \textit{epchain} and the \textit{emchain}  {\sc SAS}-tools respectively, we perform a thorough check for pile-up using the task \textit{epatplot}. The pn data do not suffer from pile-up, thus the source and the background count-rates are extracted from a circular region of radius 31.5 arcsec and 35 arcsec, respectively. Conversely, both MOS cameras appear to be slightly piled-up. Therefore, to be conservative, the final source count-rates are extracted from annuli of the same outer radius 35 arcsec and inner radius 2.75 arcsec and 2.5 arcsec for the MOS\,1 and the MOS\,2, respectively. The background count-rates are extracted from two circular regions of 3.33 arcmin and 3.42 arcmin for the MOS\,1 and the MOS\,2, respectively. Note that these apertures yield the optimum signal-to-noise ratio above 7 keV. We verify that the resulting light-curves are not affected by pile-up problems.\par
\begin{figure}
\includegraphics[width=3.65in]{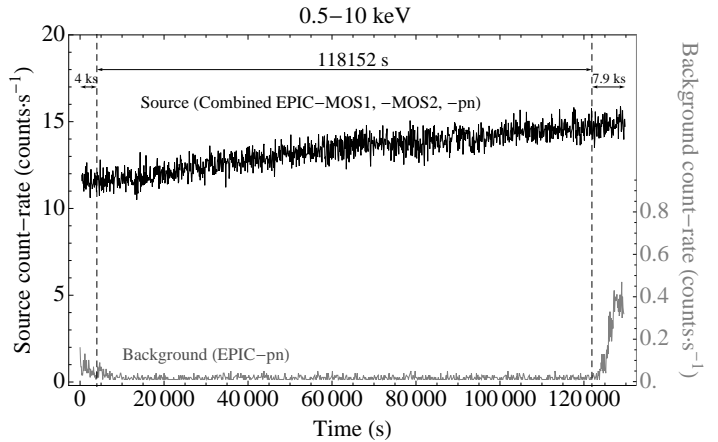}
\caption{The \f9 combined light-curve (black-points, corresponding to the left-axis) and the background pn light-curve (grey-points, corresponding to the right-axis) in the 0.5--10 keV energy-band. The background shows increased activity during the first 4 ks and the last 7.9 ks.}
\label{fig:lc_source_bkg}
\end{figure}

 
For the production of the light-curves at a given energy-range, we select events that are detected up to quadruple pixel-pattern on the CCDs i.e.\ PATTERN$<$12. The corrected background-subtracted light-curves of the source are produced using the {\sc SAS}-tool \textit{epiclccorr}. Based on the pn background light-curve, in the 0.5--10 keV energy-range (Fig.~\ref{fig:lc_source_bkg}, grey-points corresponding to the right-axis), an increased background activity was registered during the first 4 ks and the last 7.9 ks of the observations and thus, data from those periods are disregarded from our analysis\footnote{Despite the fact that the `high' background-states are still minimal, in terms of count-rate, with respect to the source count-rate in 0.5--10 keV (Fig.~\ref{fig:lc_source_bkg}, black-points corresponding to the left-axis), they consist of events having energies more than 2 keV. At these energies the source count-rate is of the order of 1.5 counts$\cdot$s$^{-1}$.}. This leaves us with 118152 s clean exposure-time.\par
For the production of the spectra we use the {\sc SAS}-task \textit{evselect} and we select for the pn camera the pulse-invariant (PI) channels from 0 to 20479, having a spectral-bin-size of 5 eV. For the two MOS cameras we select the PI channels from 0 to 11999 with a spectral-bin-size of 15 eV. For the spectral analysis we have set more stringent filtering criteria from those in the case of the light-curves. For the pn data we allow events that are detected up to double pixel-pattern on the CCD i.e.\ 
PATTERN$<$4 and we exclude all the events that are at the edge of a CCD and at the edge to a bad pixel i.e.\ FLAG$=$0. For the MOS data we allow, as in the case of the light-curves, PATTERN$<$12 but we restrict to those events having also FLAG$=$0. After filtering the events, we calculate the area of the source and the background annular regions using the task \textit{backscale} and finally we compute the detectors' response matrices and the effective areas using \textit{rmfgen} and \textit{arfgen}, respectively.\par
The X-ray spectral fitting analysis (Sect.~\ref{sect:epic_spec}) has been performed by the high-level task \textit{XSPEC} \citep{arnaud96} version 12.6.0, of the {\sc HEAsoft} version 6.10. We group the PI channels of the source-spectra, making use of the task \textit{grppha} of the {\sc FTOOLS} \citep{blackburn95} version 6.10. The MOS source-spectra are grouped in bins of four ($\Delta E$=60 eV) and eight ($\Delta E$=120 eV) spectral-channels  for the energy-ranges of 0.5--8 keV and 8--10 keV, respectively. The pn source-spectra are grouped in bins of four ($\Delta E=$20 eV) spectral-channels for the energy-range of 0.5--10 keV. All the grouped spectral-bins contain at least 20 photons.

\subsection{RGS data-reduction}
\label{ssect:rgs_obs_reduc}
The RGS raw-data are processed following the standard data-reduction threads of {\sc SAS}. We use the tool \textit{rgsproc} to extract calibrated first-order spectra and responses for the RGS\,1 and RGS\,2 cameras. In general, \textit{XMM-Newton} RGS observations can be affected by high particle background periods during parts of the \textit{XMM-Newton} orbits, mostly caused by Solar activity. The high energy-band is the one being mostly affected by background flares, and high energy RGS photons are dispersed over the CCD-9 chip. We therefore extract the background light-curve of the CCD-9 chip and select, as good-time-intervals of the processed observation, only those during which the background count-rate deviates by less than two standard deviations from the average background count-rate of each observation. The final (i.e.\ after cleaning for high-background time-intervals) exposures of the RGS\,1 and RGS\,2 spectra are 129.6 ks and 129.8 ks, respectively.\par Both RGS\,1 and RGS\,2 spectra are grouped at a resolution of 8 m\AA\ ($\sim$0.16 eV at 0.5 keV), so allowing 8 spectral-bins for each RGS resolution element. For both the RGS\,1 and the RGS\,2, the photon count-rate of the background spectrum becomes comparable to the background-subtracted source count-rate, at energies less than 0.35 keV ($\lambda\ge35$ \AA), and greater than 1.8 keV ($\lambda\le$7 \AA). Moreover, due to failures of two different read-out detector chips, early in the life of the mission, both the RGS\,1 and the RGS\,2 lack response in two different spectral intervals of 0.9--1.2 keV and 0.5--0.7 keV, respectively. We therefore consider the following two spectral intervals for spectral fitting purposes: $[(0.35-0.9) \bigcup (1.2-1.8)]$ keV and $[(0.35-0.5) \bigcup (0.7-1.8)]$ keV, for RGS\,1 and RGS\,2, respectively.\par
For the spectral fits the \textit{Sherpa} modelling and fitting application of the {\sc Chandra Interactive Analysis of Observation} ({\sc CIAO}) software version 4.2, has been used to fit simultaneously the RGS\,1 and RGS\,2 spectra of \f9. In addition, the spectral fitting analysis results have been cross-checked and verified with \textit{XSPEC}, but the latter values are less conservative and thus we report only the results coming from \textit{Sherpa}.   

\section{X-RAY LIGHT-CURVES}
\label{sect:lcs}
\begin{figure}
\hspace{0.26em}\includegraphics[width=3.444in]{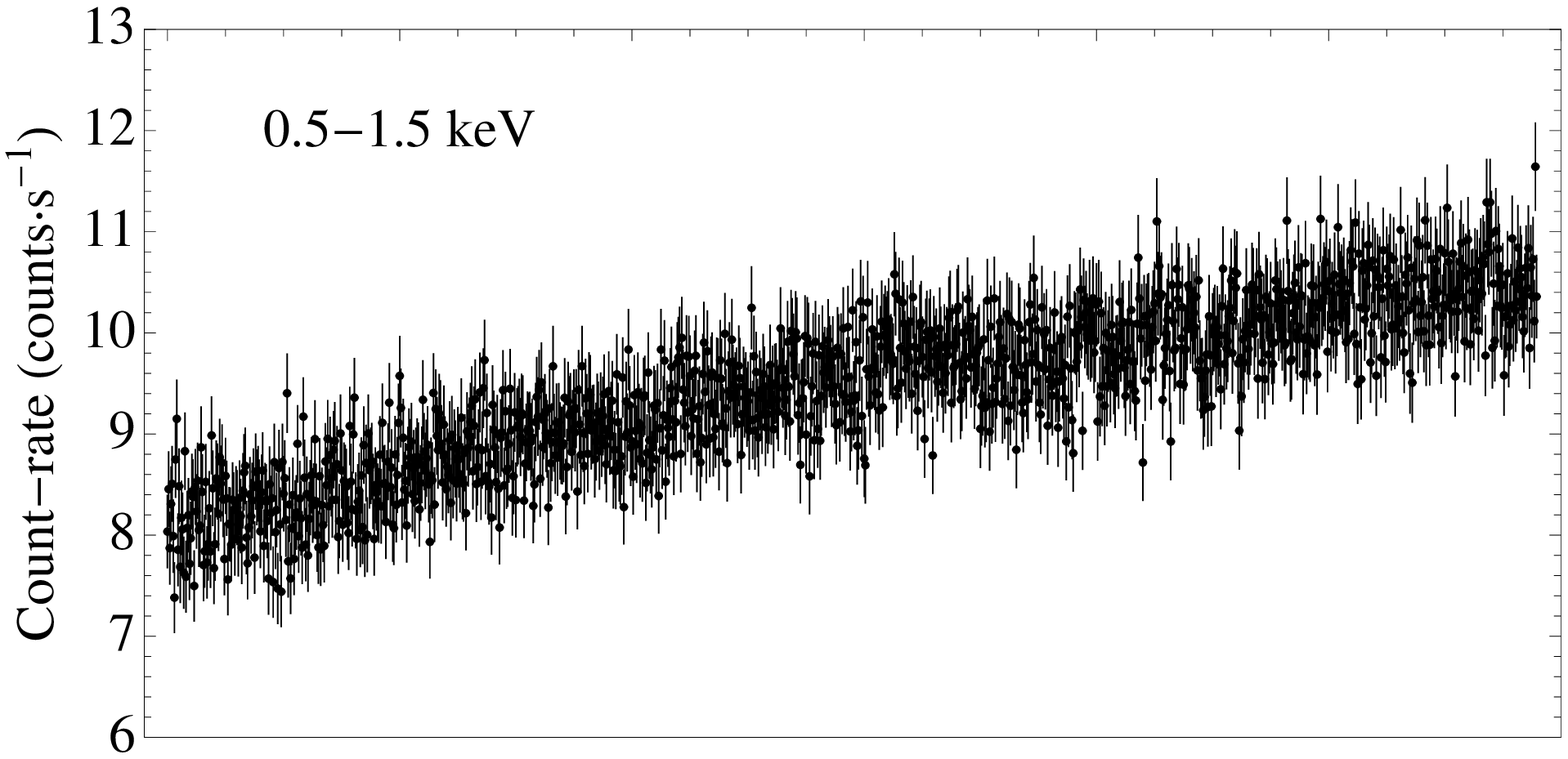}\\[-1.8em]
\hspace*{-0.03em}\includegraphics[width=3.475in]{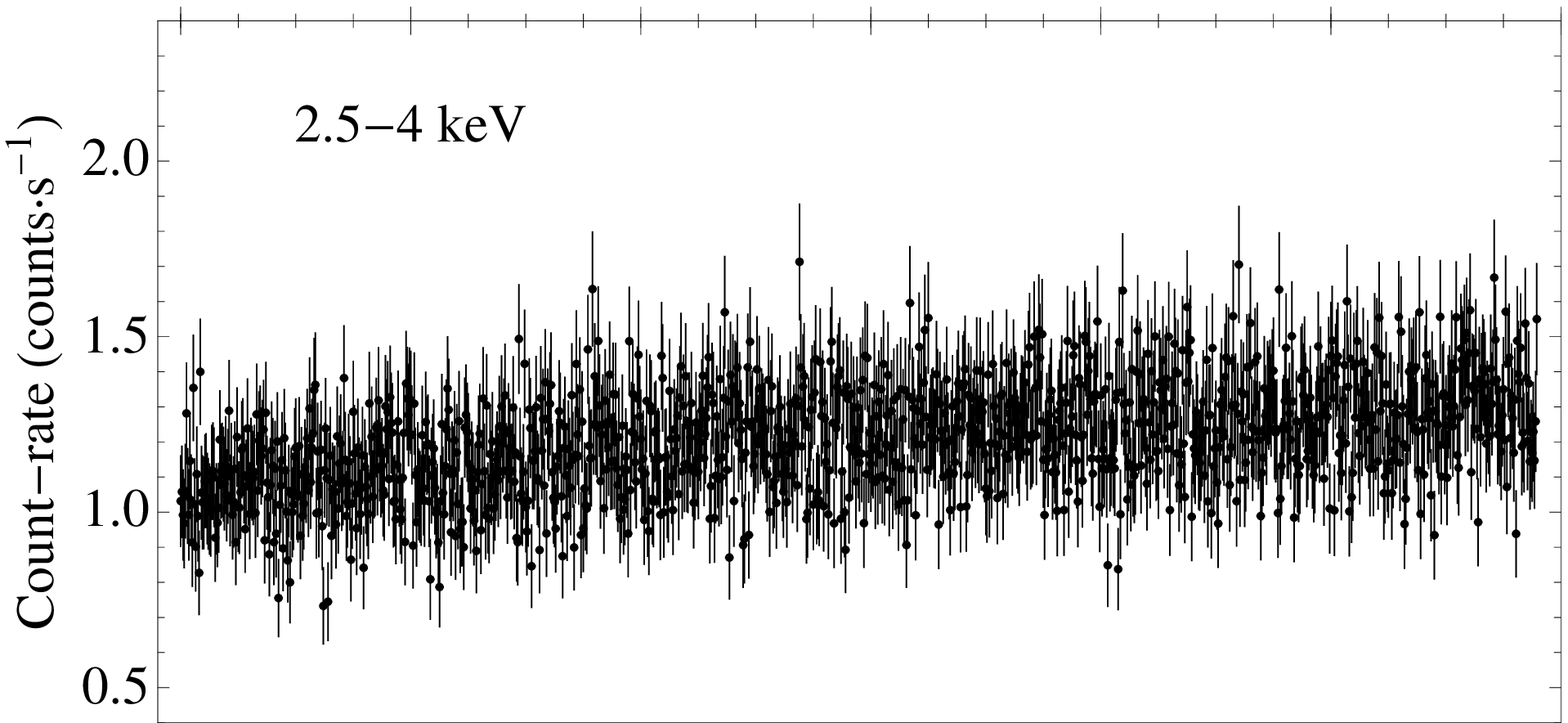}\\[-1.42em]
\hspace*{-0.22em}\includegraphics[width=3.699in]{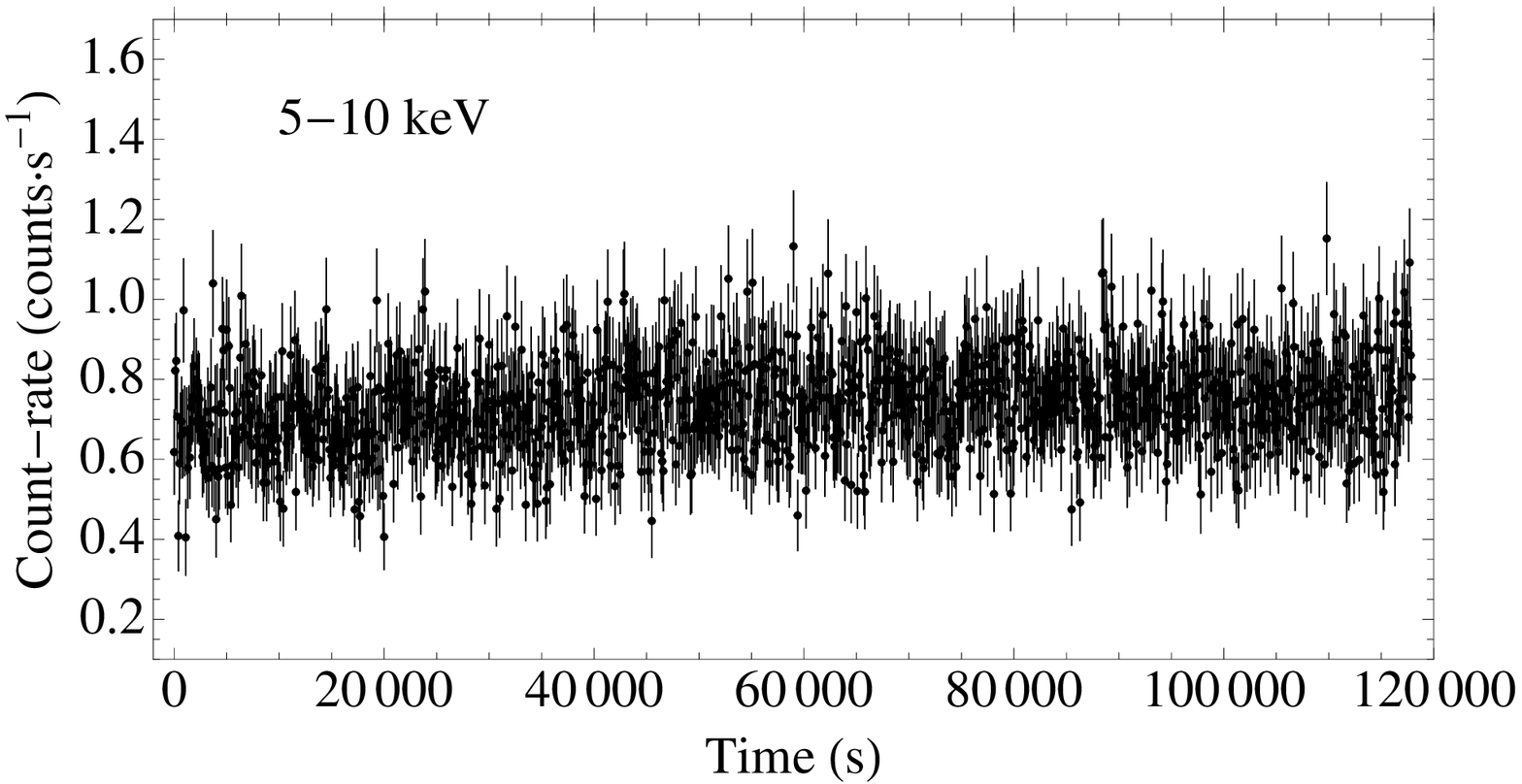}
\caption{The combined EPIC-MOS\,1,-MOS\,2,-pn X-ray light-curves of \f9 in the 0.5--1.5 keV, 2.5--4 keV, and 5--10 keV energy-bands, in bins of 100 s. Note the difference in the count-rate scale on the vertical axis.}
\label{fig:lightcurves}
\end{figure}

The combined X-ray light-curves from the EPIC-pn and the two EPIC-MOS cameras in 0.5--1.5 keV, 2.5--4 keV, and 5--10 keV energy-ranges, are shown in Fig.~\ref{fig:lightcurves}. A steady count-rate increase can be noticed in all three energy-bands, of the order of 30 per cent, 20 per cent, and 10 per cent, respectively. Additionally, the fractional variability amplitudes \citep[corrected for the photon noise e.g.][]{vaughan03}, for the three bands are 7.3$\pm$0.1 per cent, 5.7$\pm$0.6 per cent, and 2.2$\pm$2.2 per cent, respectively. Both the overall count-rate and the fractional variability amplitude increase with decreasing energies.

\subsection{Hardness-ratio analysis}
\label{ssect:hrs}
\begin{figure}
\includegraphics[width=3.5in]{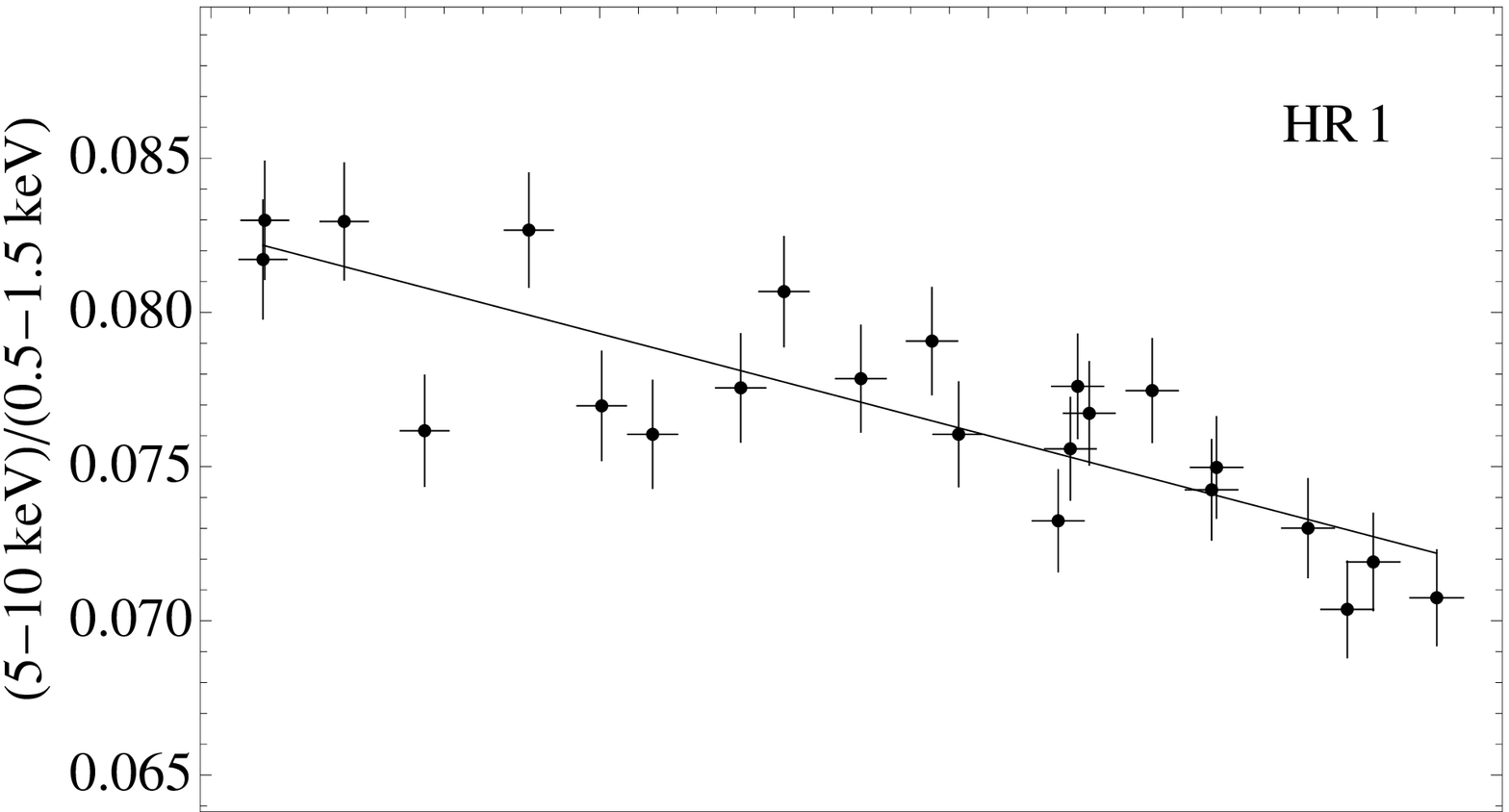}\\[-1.43em]
\hspace*{0.63em}\includegraphics[width=3.427in]{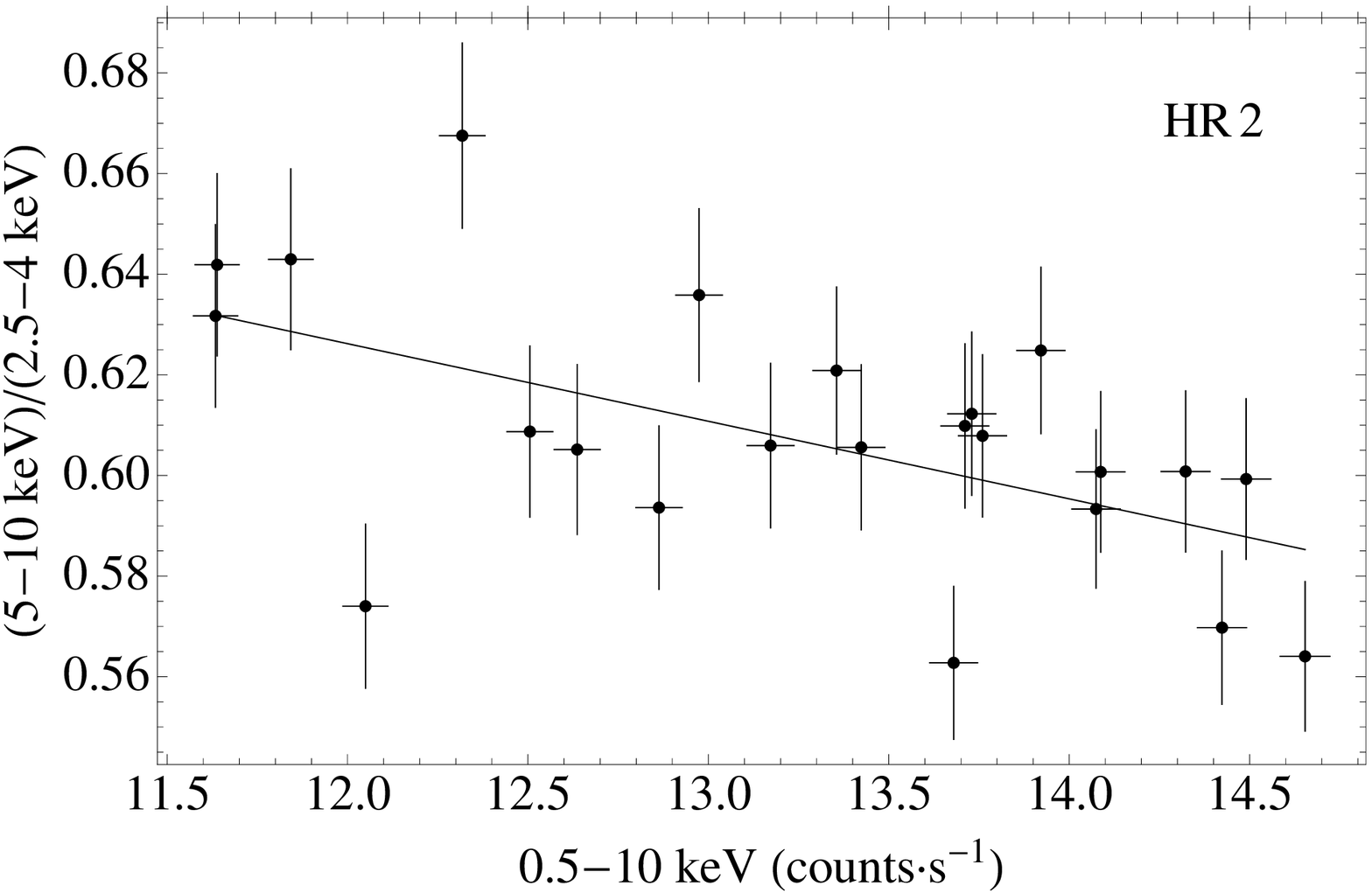}
\caption{Hardness-ratios in (5--10 keV)$/$(0.5--1.5 keV) and (5--10 keV)$/$(2.5--4 keV) versus the total count-rate in 0.5--10 keV, binned in 5 ks time-bins. The solid-line represents the best-fit linear model. The negative slope implies a `softer-when-brighter' X-ray spectral behavior.}
\label{fig:hr_plot}
\end{figure}

Observing different variability amplitudes in different energy-bands is usually indicative of spectral variations \citep[e.g.][]{brinkmann07}. After binning the light-curves in bins of 5 ks, we estimate the hardness-ratios (5--10 keV)$/$(0.5--1.5 keV) (HR\,1) and (5--10 keV)$/$(2.5--4 keV) (HR\,2) versus the overall count-rate in 0.5--10 keV (Fig.~\ref{fig:hr_plot}). We see a decreasing trend implying that the source X-ray spectrum becomes softer when the source gets brighter.\par
In order to check the significance of such a trend, we fit to the 23 hardness-ratio points a linear model and we check how significantly the value of the slope differs from that of zero. For HR\,1 the best-fit-model gives a slope of $(-3.3\pm0.5)\times10^{-3}$ with a 95 per cent confidence interval (c.i.) of ($-4.3\times10^{-3}$,$-2.3\times10^{-3}$)\footnote{The value of ${\itl t}$-statistic is ${\itl t}_{21,0.025}=2.08$}. Since the value of ${\itl t}$-statistic that we get from the data-set is 18.71, the probability of getting such a value by chance alone is only $1.42\times10^{-14}$.\par
For HR\,2 the best-fit model gives a slope of $(-1.5\pm0.5)\times10^{-2}$ having a 95 per cent c.i. of ($-2.6\times10^{-2},-0.5\times10^{-2}$). The value of ${\itl t}$-statistic for this data-set is 11.95 yielding a chance coincidence probability of only $7.86\times10^{-11}$. Since in both cases the zero value lies outside the estimated c.i., the null-hypothesis (i.e.\ the slope is equal to zero) is rejected at 5 per cent significance level.\par
This result implies that \f9 shows significant X-ray spectral variations, becoming `softer-when-brighter'. This behavior is considered typical for radio-quiet type $\rmn{I}$ Seyfert \citep[e.g.][]{sobolewska09}.

\section{THE RGS SPECTRUM OF FAIRALL\,9}
\label{sect:rgs_spec}
\begin{figure}
\includegraphics[width=84mm]{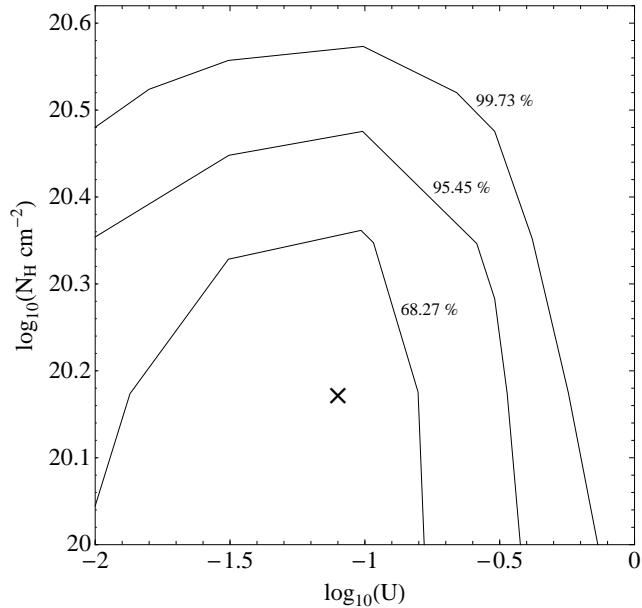}
\caption{The 68.27 per cent, 95.45 per cent, and 99.73 per cent confidence contours for the best-fit parameters, $\log_{\rm 10}\left(U\right)$ and $\log_{10}\left(N_{\rm H}\,{\rm cm^{-2}}\right)$ indicated by the cross, for the best-fit power-law plus warm-absorber model to the RGS\,1 and RGS\,2 spectra.}
\label{fig:wa_contour}
\end{figure}

We search the \f9 RGS\,1 and RGS\,2 spectra (in their available wavelength intervals) for narrow absorption and emission features. We inspect locally 2 \AA\ wide spectral regions, and we model the local continua with simple power-laws attenuated, in this line of sight, by the Galactic neutral absorption $N_{\rm H}=3.2\times10^{20}$cm$^{-2}$ \citep[estimated using the {\sc FTOOL} \textit{nH}, after][]{dickey90}. When both the RGS\,1 and the RGS\,2 are available we fit simultaneously the same 2 \AA\ regions from the two spectrometers. In Fig.~\ref{fig:wa_contour} we show the 18--24 \AA\ portion of the RGS\,1 spectrum of \f9, together with its best-fit continuum model (red-line) and the corresponding ratio plot (data$/$model). Note that in this section all the centroid-energies of the emission- absorption-lines refer always to the observer's frame.\par
From the ratio plot of Fig.~\ref{fig:wa_contour} we find no significant evidence for absorption-lines imprinted either by intrinsic (i.e.\ nuclear) or by intervening (i.e.\ interstellar medium (ISM) or intergalactic medium) material down to 4 standard deviations (below the fluctuations of the continuum) line equivalent-width (eqw) sensitivity of 45 m\AA. The only exception is a relatively strong local O\,{\sc i}\,K$\alpha$ line (eqw$=36^{+25}_{-11}$ m\AA), imprinted by our Galaxy ISM at E=$0.5273^{+0.0004}_{-0.0013}$ keV (23.51$^{+0.02}_{-0.06}$ \AA), and detected at a confidence level of 2.9 standard deviations (below the fluctuations of the continuum). Assuming a relative fraction (with respect to the ensemble of the oxygen species) of O\,{\sc i} of the order of unity, i.e.\ neutral material, this gives an average ISM metallicity along the line of sight of the order of 20 per cent that of the Solar value. Moreover, we find evidence of two additional local absorption-lines O\,{\sc vii}\,K$\beta$ (eqw$=39^{+15}_{-10}$ m\AA) and  O\,{\sc vii}\,K$\alpha$ (eqw$=38^{+17}_{-12}$ m\AA) at E=$0.6640^{+0.0003}_{-0.0013}$ keV (18.67$^{+0.02}_{-0.04}$ \AA) and E=$0.5746^{+0.0005}_{-0.0013}$ keV (21.58$^{+0.02}_{-0.05}$ \AA), respectively. These lines are detected at a confidence level of 3.8 and 3.2 standard deviations (below the fluctuations of the continuum) and they are thought to be created by an ubiquitous local absorber (seen along virtually all lines of sight), which could be hot gas in an extended corona of our galaxy, or a local warm-hot intergalactic medium filament permeating our own Local Group and/or Milky Way, or both \citep{nicastro02}. However, from the ratio plot of Fig.~\ref{fig:wa_contour} (dashed-lines at 18.67 \AA\ and 21.58 \AA) we notice that each of these two absorption-lines coincides with effective area features, i.e.\ instrumental features.\par
\begin{figure}
\includegraphics[width=83mm]{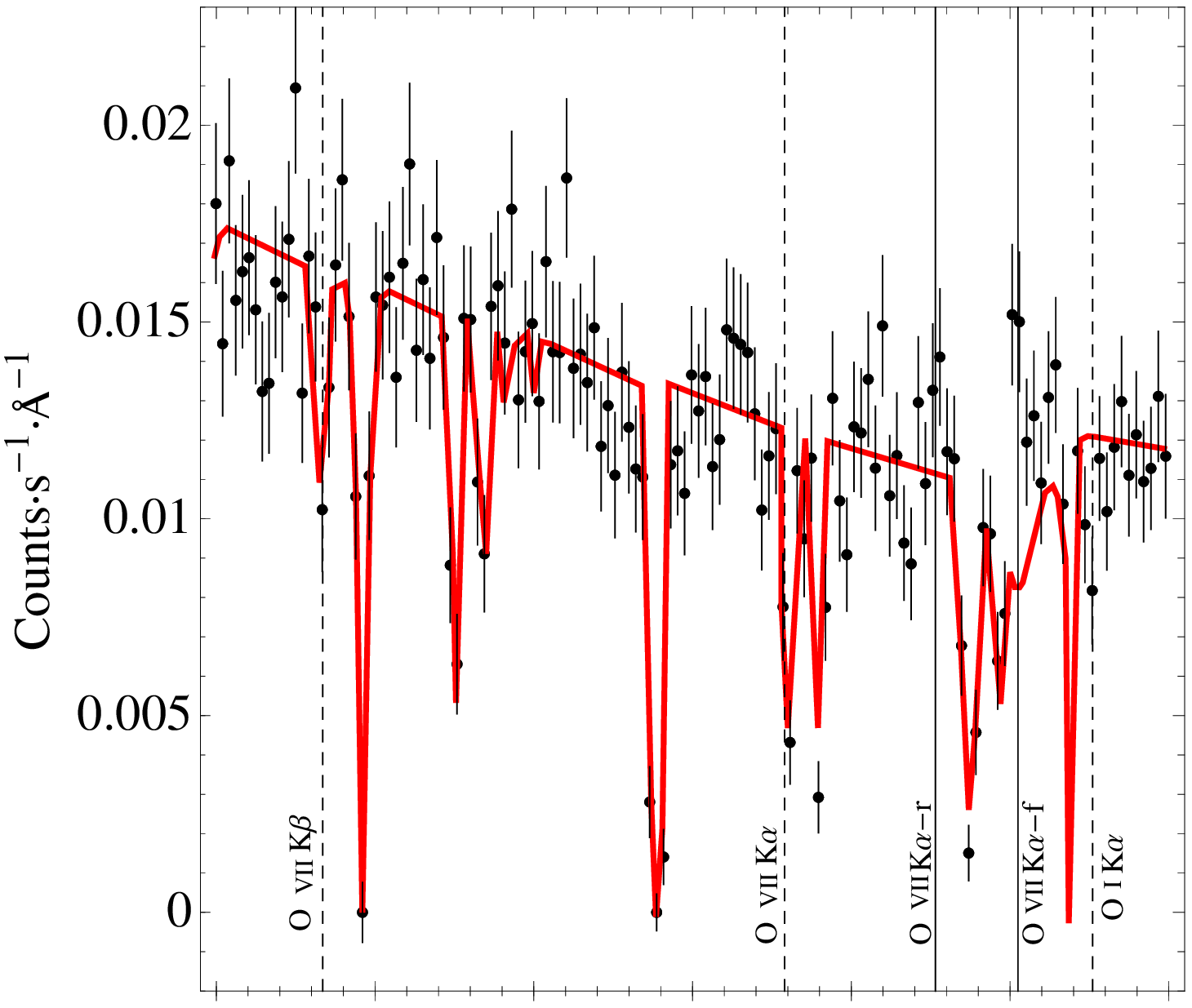}\\[-1.63em]
\hspace*{1.91em}\includegraphics[width=77.7mm]{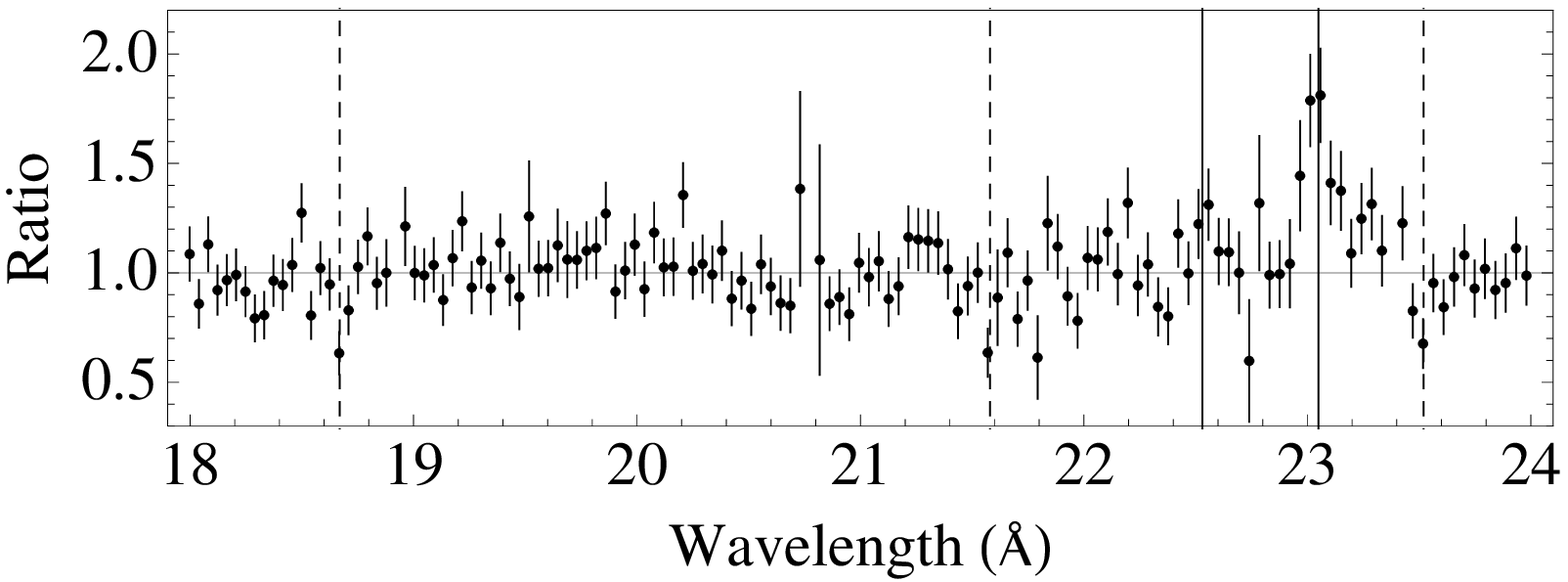}
\caption{Spectral fit in the 18--24 \AA\ wavelength-band of the unbinned (i.e.\ $\Delta\lambda=8$ m\AA) RGS\,1 spectrum. The spectral model is a power-law attenuated Galactic neutral absorption (red-line). Notice that the deep absorption-like features, appearing in this spectral interval, are instrumental features of the RGS\,1 dispersion CCD detectors (i.e.\ dead pixel/columns), and are modelled by the RGS\,1 effective area. The only real absorptions features are the weak absorptions-lines (dashed-lines) created locally (i.e.\  Milky Way and/or in the Local Group) O\,{\sc i}\,K$\alpha$, O\,{\sc vii}\,K$\beta$, and O\,{\sc vii}\,K$\alpha$. The only real emission features present are the O\,{\sc vii}\,K$\alpha$-$r$ and O\,{\sc vii}\,K$\alpha$-$f$ emission-lines (solid-lines) detected at the systematic redshift of \f9.}
\label{fig:rgs_spec}
\end{figure}

Additionally, we find evidence for the presence of few narrow (i.e.\ unresolved) emission-lines close to the systemic redshift of \f9. In particular, we identify two of the three lines of the O\,{\sc vii}\,K$\alpha$ emission triplet, the resonant (r) and the forbidden (f) lines (Fig.~\ref{fig:rgs_spec}, solid-lines). During the fit, the three lines (r,f and intercombination (i)) have positions linked to their rest-frame line energy-ratio and only the energy of O\,{\sc vii}\,K$\alpha$-r is left to vary freely. The rest-frame wavelengths of these lines are 21.602 \AA\ (r), 21.8 \AA\ (f), and 22.1 \AA\ (i). Unfortunately, the O\,{\sc vii}\,K$\alpha$-i falls into an instrumental feature of the RGS\,1 (the RGS\,2 is blind at these wavelengths) and thus no density diagnostics are possible for the X-ray emitting gas. The O\,{\sc vii}\,K$\alpha$-f line (eqw$=180^{+83}_{-39}$ m\AA) at E=0.5379 keV (23.05 \AA) and O\,{\sc vii}\,K$\alpha$-r line (eqw=42$^{+35}_{-22}$ m \AA) at E=$0.5502\pm0.0004$ keV (22.53$\pm0.02$ \AA) are detected at a confidence level of 4.5 and 1.9 standard deviations (above the fluctuations of the continuum) at a redshift of z=0.043$\pm$0.001, implying an outflow velocity relative to the optical systemic redshift of \f9 of -1200$\pm 300$ km s$^{-1}$. The O\,{\sc vii}\,K$\alpha$ (f/r) 4.3$^{+4.1}_{-2.4}$ intensity ratio strongly suggests the photo-ionization nature of the emitting gas \citep{porquet00}.\par
Despite the absence of individually significant absorption-lines, we check further for the presence of any ionized gas in the source. To this end, we jointly fit the RGS\,1 and RGS\,2 spectra with a model consisting of a power-law plus a warm-absorber attenuated, in this line of sight, by the Galactic neutral absorption $N_{\rm H}$. We use the self-consistent code {\tt Photoionized Absorption Spectral Engine} ({\tt PHASE}) \citep{krongold03}, that calculates the absorption caused by ionized plasma. The code has four parameters: the equivalent hydrogen column density of the absorber, the ionization parameter \textit{U} of the gas (defined as the ratio between the density of ionizing photons and the electron density of the gas), the turbulent velocity, and the redshift of the absorber.\par
We freeze the absorber's redshift to the systemic redshift of \f9, and its turbulent velocity to v$_{\rm turb}=100$ km s$^{-1}$, while we leave the equivalent hydrogen column density and the ionization parameter $U$ of the absorber free to vary during the fit. The best-fit value that we obtain for $\log_{\rm 10}\left(U\right)$ is $-1.1_{-0.6}^{+0.2}$ is rather low for typical type $\rmn{I}$ Seyfert warm-absorbers. Furthermore, although we can obtain just an upper limit for the equivalent hydrogen column-density of the absorber, this limit is rather tight: $\log_{10}\left(N_{\rm H}\,{\rm cm^{-2}}\right) \le 20.36$ (the errors correspond to one standard deviation for the two interesting parameters). In Fig.~\ref{fig:wa_contour} we show the 68.27, 95.45 and 99.73 per cent confidence contours for these two parameters. Clearly, a significant amount of intrinsic absorbing material along our line of sight to \f9 is not required by the RGS data.

\section{EPIC X-RAY SPECTRAL ANALYSIS}
\label{sect:epic_spec}
For the X-ray spectral results, all the errors on the best-fit parameters indicate their 90 per cent confidence range, corresponding to a $\Delta \chi^2$ of 2.706, unless otherwise stated. During the fitting procedures, we keep all the model-component parameters between the two MOS and the PN spectra tied together except from the normalizations (for the two MOS spectra they are still tied together). The model-flux is modified using the photoelectric absorption model {\tt wabs} having a fixed interstellar column-density of $N_{\rm H}$. Note, that the best-fit values for the centroid energy of the emission-lines are given in the source's rest-frame and the horizontal axis in the spectral-plots always refers to the energies in the observer's frame.

\subsection{A first look at the broad-band spectrum (Model 1)}
\begin{figure}
\includegraphics[width=3.5in]{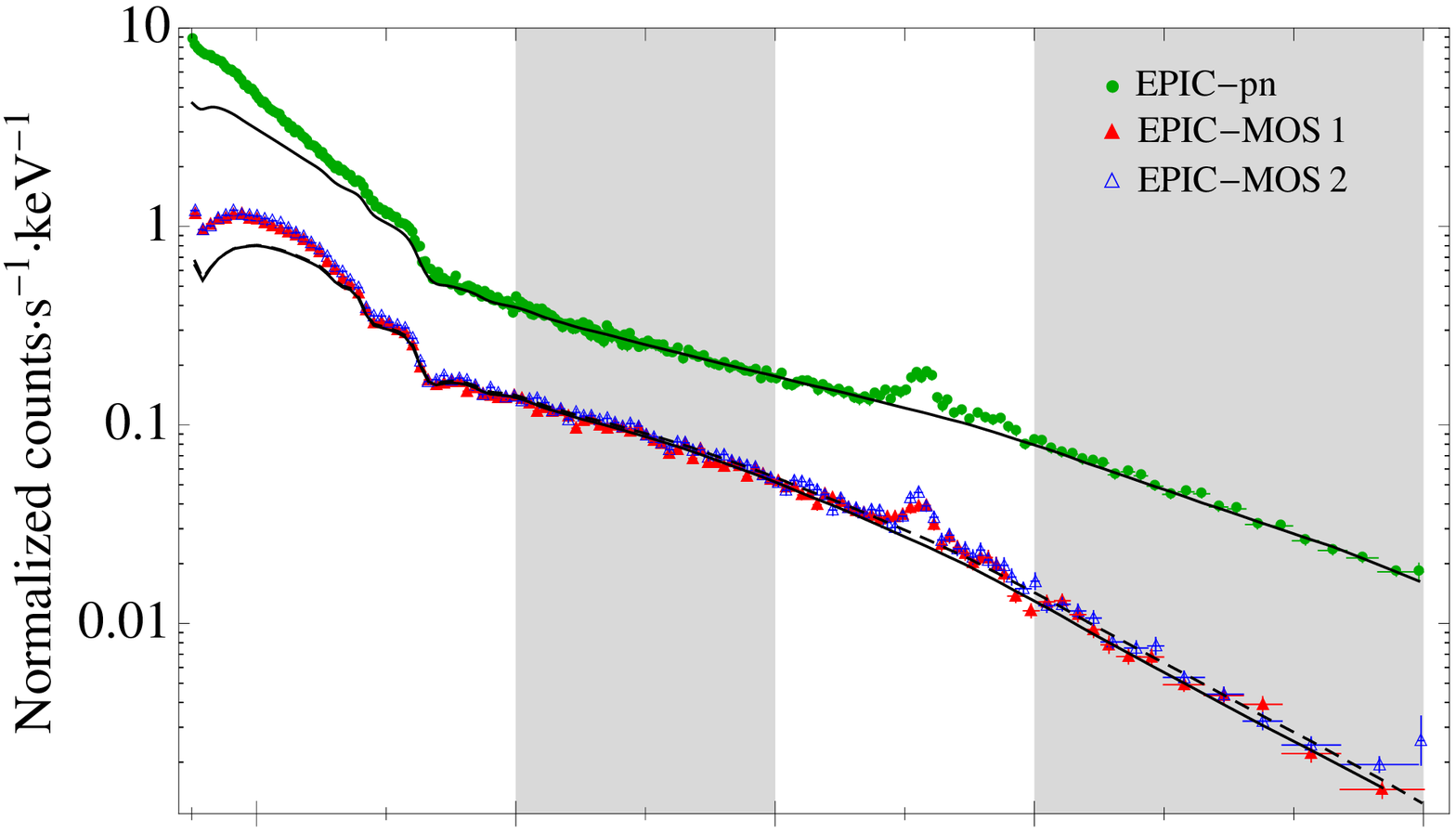}\\[-2.57em]
\hspace*{0.8em}\includegraphics[width=3.387in]{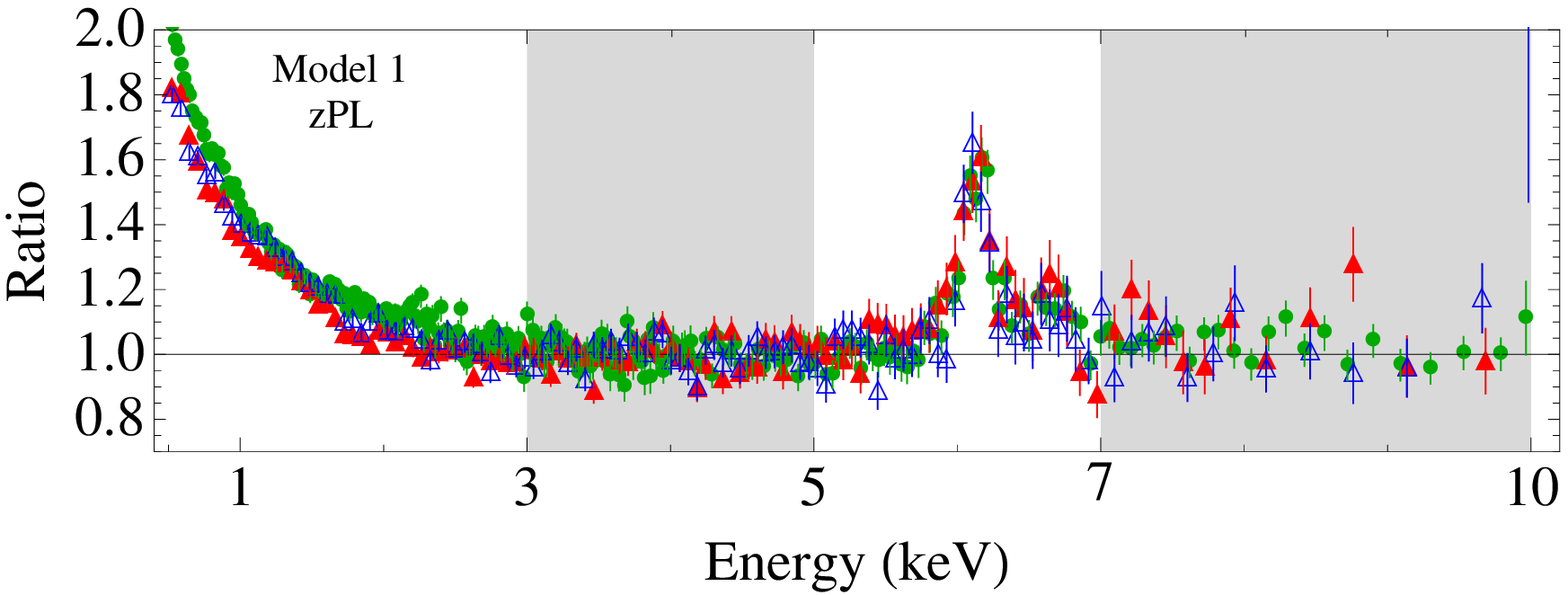}\\[-1.81em]
\caption{Model 1: Redshifted power-law fit in the 3--5 keV and 7--10 keV energy-ranges, indicated by the grey-areas. The best-fit model is extrapolated over the energy-range of 0.5--10 keV}
\label{fig:spectral1}
\end{figure}

In order to get a phenomenological view of the overall spectra we initially fit a redshifted power-law (Model 1) to the data in the 3--5 keV and 7--10 keV energy-bands. The $\chi^2$ value of the best-fit model is equal to 419.41 for 388 d.o.f. yielding a best-fit photon spectral index of $\Gamma=1.66\pm0.02$. Figure~\ref{fig:spectral1} shows this best-fit model extrapolated over the energy-range of 0.5--10 keV together with the respective ratio plot.

From the ratio-plot of Fig.~\ref{fig:spectral1} we can readily distinguish two very distinct spectral features. Below 2 keV, we notice a strong flux-excess on the top of the best-fit power-law extrapolation. Then, in the 5.7--8 keV energy-range a strong emission-line appears around 6.1 keV followed by an excess around 6.7 keV. In addition to those features, significant broad residuals appear above 7 keV as well as a feature at 6.8 keV suggestive of an absorption edge.

\subsection{Spectral fitting in the 3--10 keV energy-range }
\label{ssect:epic_spec_3_10}
Since Model 1 (Fig.~\ref{fig:spectral1}) gives an adequate description of the data in the 3--5 keV and 7--10 keV energy-ranges, we keep it as a baseline model, for the following spectral-analysis over the 3--10 keV energy-range. The best-fit parameters for the following X-ray spectral models are given in Table~\ref{tab:spec_modelFits}.
\subsubsection{Iron line and reflection (Model 2)}
\label{ssect:iron_line_reflection}
To account for the emission-line at 6.1 keV (in the observer's frame), we consider a redshifted zero-width Gaussian-function model. Following both \citet{gondoin01} and \citet{schmoll09} and due to the fact that this line appears to correspond to the \fa line from neutral iron, having a centroid energy of 6.4 keV in the source's rest-frame, we add further a redshifted zero-width Gaussian-function model to account for the associated \fb line\footnote{The exact energy position of these two iron lines depends on the ionization state of the Fe element in the source. Assuming neutral atoms \citep[after][]{kaastra93} the \fa is a doublet composed of a 6.391 keV and a 6.405 keV line, as well as the \fb line consisting of a 7.057 keV and a 7.058 keV line. Since the energy resolution of the EPIC cameras at 6--7 keV energy-range is around 150 eV (full width half maximum), it is impossible to resolve them and thus we fit them with a zero-width redshifted Gaussian-function.}. For the latter, we fixed the centroid energy at 7.06 keV (in the source's rest-frame) and its normalization at a value equal to 0.16 of the \fa normalization \citep{molendi03}.\par
We model the absorption-edge and the associated reflection-continuum as the reflection of an exponentially cut-off power-law spectrum from a neutral disc using the \textit{XSPEC} model {\tt pexrav} \citep{magdziarz95}. In order to derive the reflected continuum only, we restrict the range of the relative-reflection scaling-factor to negative values. The energy cut-off is fixed at $E_{\rm c}$=200 keV (in the source's rest-frame), the abundance of elements heavier than helium is set equal to that of the Solar value, the iron abundance is set equal 0.8 of the Solar value \citep{schmoll09}, and the inclination angle is fixed at $\phi=40$ degrees. Note that we do not tie together the photon index of the {\tt pexrav} model with that of the redshifted power-law. If the reflecting material is associated with the distant molecular torus, which is expected in Seyfert galaxies, it will probably respond to the average X-ray source-spectrum, whose slope may be different than the spectral slope during this \textit{XMM-Newton} observation.\par
The best-fit model (Model 2) is plotted in the top panel of Fig.~\ref{fig:spectral2_3}, having a $\chi^2$ value of 600.29 for 538 d.o.f. The best-fit centroid energy of the \fa line is 6.42$\pm0.01$ keV. The best-fit values for the relative reflection and the power-law photon index, from the {\tt pexrav} model, are $R{\rm pex}=0.87^{+0.05}_{-0.06}$ and $\Gamma{\rm pex}=1.87\pm0.03$, respectively. Finally, the best-fit power-law photon index, coming from the redshifted power-law model-component is $\Gamma=1.85^{+0.03}_{-0.02}$. From the ratio plot of the top panel of Fig.~\ref{fig:spectral2_3}, we see that even after including the two narrow Fe lines and the reflection component, three broad emission components still remain around 5.9 keV, 6.1--6.7 keV, and 6.9--7.7 keV (in the observer's frame).\par
Note that the photon indices coming from the {\tt pexrav} and the power-law model-components are consistent within their errors. Tying the two photon-indices together and repeating the fit yields a value of $\Gamma=1.87^{+0.01}_{-0.02}$, being consistent with the previously derived values.
  
\subsubsection{Adding a relativistic disc-line (Model 3)}
\label{ssect:epic_rel_disc_line}
In order to quantify further the aforementioned spectral-excesses, we fix the parameters of Model 2 to their best-fitted values and we add the relativistic disc-line model of \citet{laor91}, denoted as {\tt laor} in \textit{XSPEC}, (Model 3). The latter model-component describes the line-profile emitted from an accretion disc around a rotating black hole including the strong relativistic effects that are produced in the innermost regions of the accretion disc.\par
\begin{figure}
\hspace*{0.01em}\includegraphics[width=3.503in]{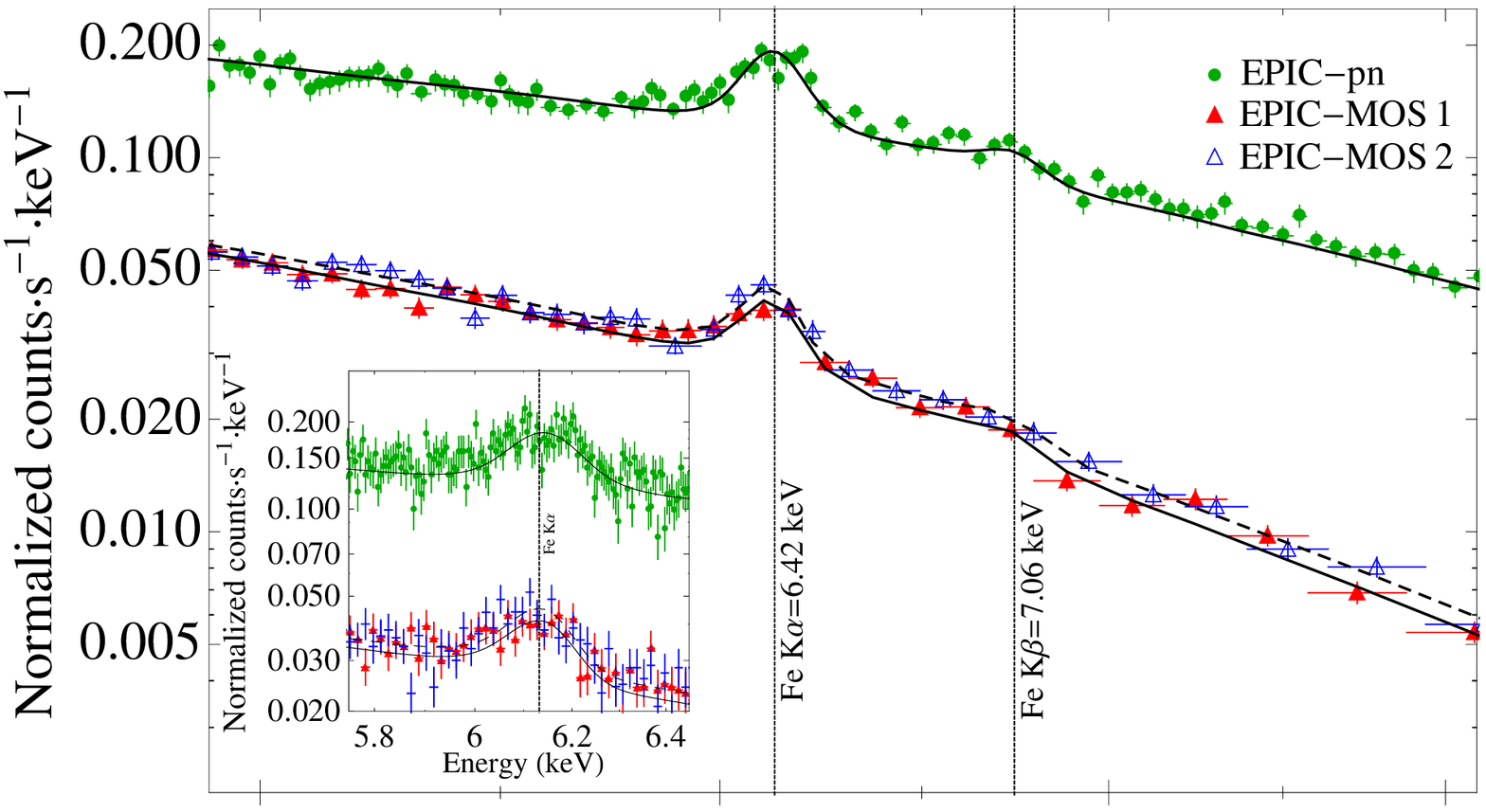}\\[-1.43em]
\hspace*{1.415em}\includegraphics[width=3.317in]{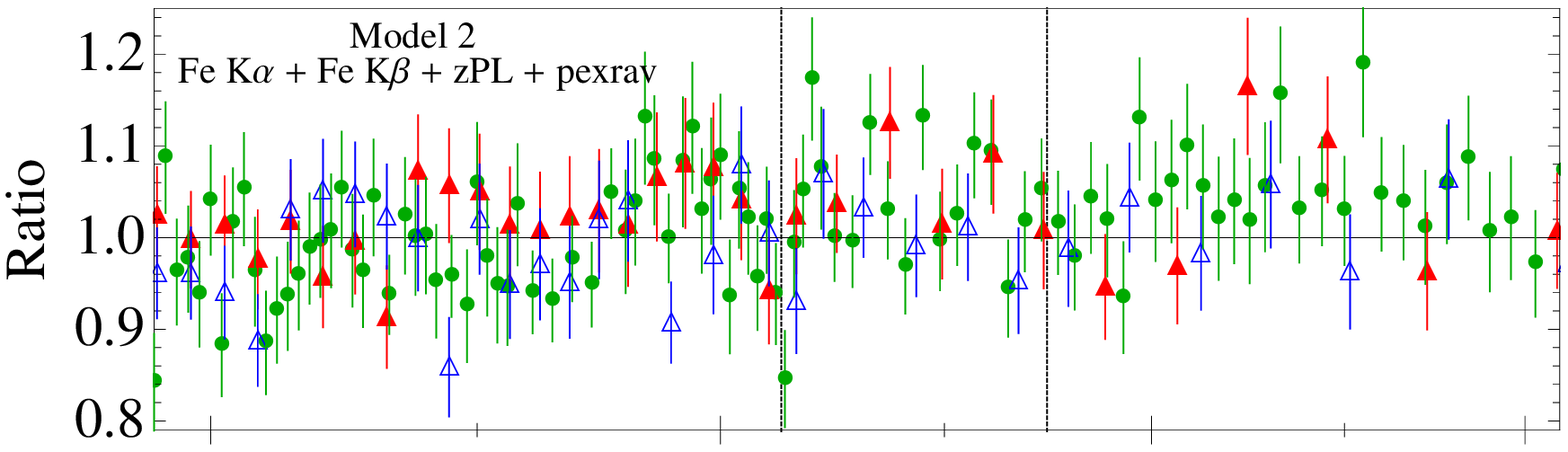}\\[-2.43em]
\hspace*{0.622em}\includegraphics[width=3.429in]{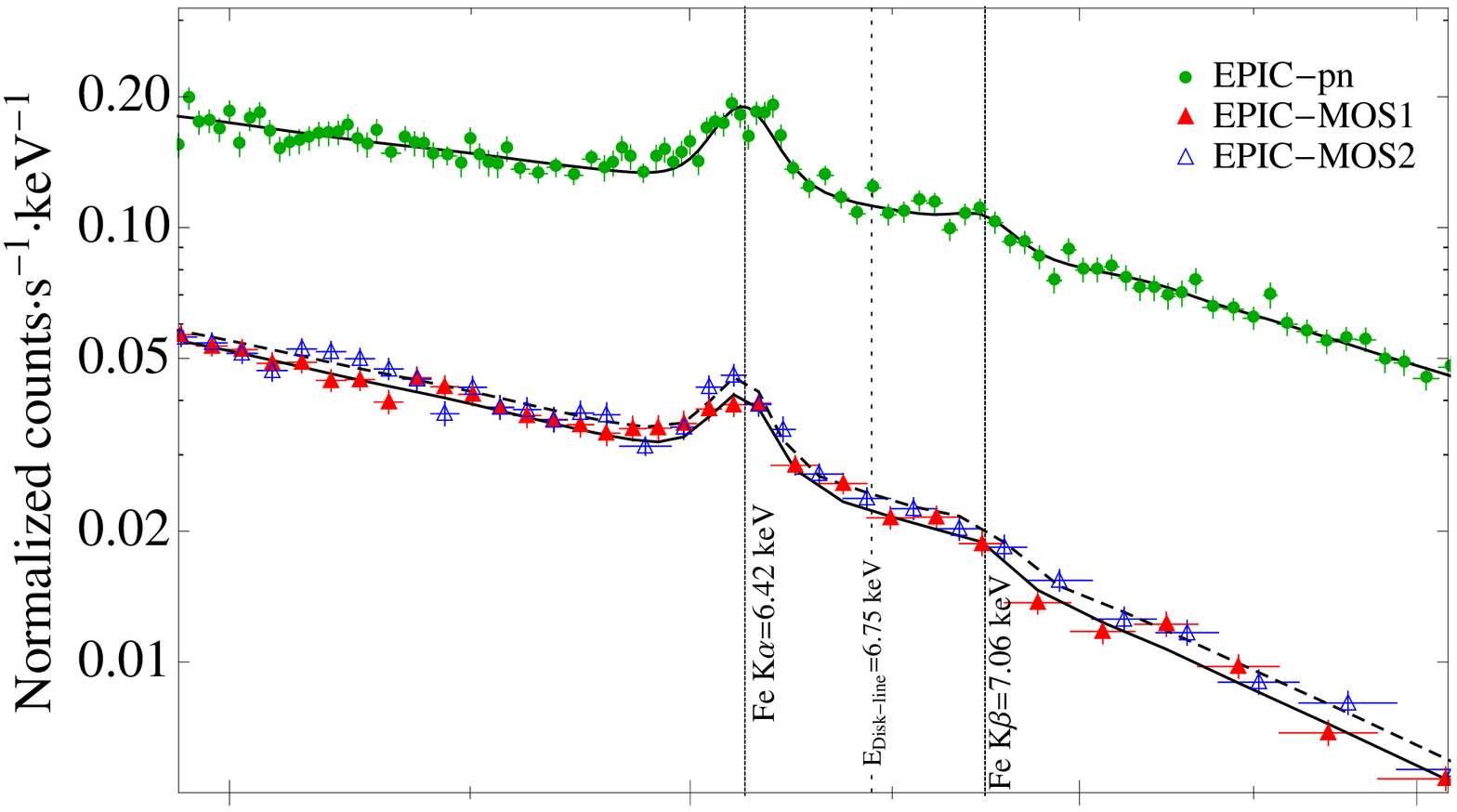}\\[-2.51em]
\hspace*{1.41em}\includegraphics[width=3.318in]{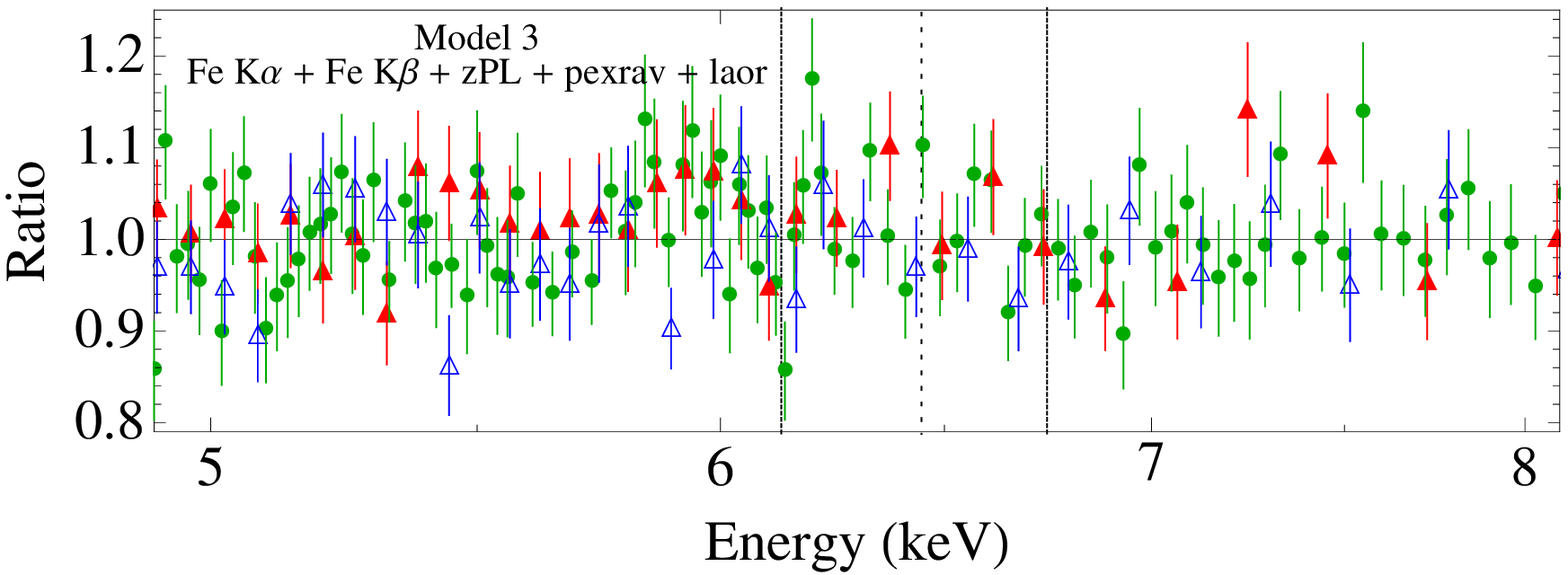}
\caption{Spectral fits in the 3--10 keV energy-range. The vertical labels indicate the energy in the source's rest-frame. (Top panel) Model 2: Two narrow Fe lines --\fa (free), \fb (fixed)--, a redshifted power-law, and a reflection component. The inset shows the model together with the actual spectral channels of the pn and MOS detectors, separated by 10 eV and 15 eV, respectively. (Lower panel) Model 3: Two narrow Fe lines --\fa (fixed), \fb (fixed) --, a redshifted power-law, a reflection component (fixed), and a relativistic disc-line.}
\label{fig:spectral2_3}
\end{figure}
\begin{figure}
\includegraphics[width=3.5in]{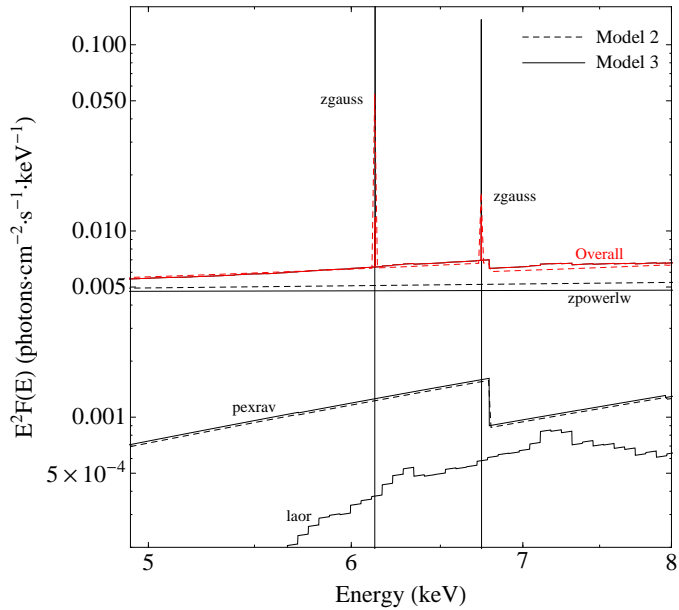}
\caption{Spectral components for the best-fit Model 2 (black dashed-lines) and best-fit Model 3 (black solid-lines) to the pn data. The red lines correspond to the overall best-fit theoretical models respectively.}
\label{fig:components2_3}
\end{figure}

During the fit, we fix the parameters of the inner and the outer radius of the accretion disc to 6 and 300 gravitational radii, $r_{\rm g}$, respectively. From the bottom panel of Fig.~\ref{fig:spectral2_3} we can see that Model 3 gives a very good description of the excesses in the 6.1--7.7 keV energy-range (in the observer's frame). The $\chi^2$ value of this fit is 559.18 for 536 d.o.f. The best-fit value for the centroid energy of the relativistic line is 6.75$\pm 0.18$ keV. Also, the best-fit values for the emissivity index and the inclination angle are $q=2.46^{+0.40}_{-0.37}$ and $\theta=84^{+2}_{-17}$ degrees, respectively. Finally, the power-law photon index is steeper than that of Model 2 and equal to $\Gamma=1.97^{+0.07}_{-0.06}$. In Fig.~\ref{fig:components2_3} we show the spectral components of both the best-fit Model 2 (dashed-line) (Sect.~\ref{ssect:iron_line_reflection}) and the best-fit Model 3 (solid-line) to the pn data, depicting the relative flux contribution of each component to the X-ray spectrum.\par
For this model, we can now estimate the eqw of the narrow and broad emission-lines, having as an underlying-continuum the contribution from all other components of Model 3. The eqws for the \fa and \fb lines that we get are 103$\pm19$ eV and 19$\pm4$ eV, respectively (mean values from the pn and the MOS data). The eqw of the relativistic disc-line is 362$^{+218}_{-305}$ eV and 162$^{+254}_{-162}$ eV, for the pn and the MOS data, respectively.\par
Theoretically, it is expected that the eqw of the narrow emission-lines should be tightly connected with the amount of the continuum reflection component \citep{matt91}. In fact, one can consider a linear correlation between the two parameters with the reflection strength, $R=\frac{\Omega}{2\pi}$, ($\Omega$ is solid angle that the reflector subtends) being equal to unity when the \fa line's eqw is 120 eV \citep[e.g.][]{matt91,bianchi09}. For the best-fit eqw-value of \fa line, that we have found previously, the expected reflection should be $R=0.86\pm 0.17$. This value is entirely consistent with the reflection fraction estimates mentioned above (Model 2). Based on the estimated reflection value, the average solid angle that the distant reflector should subtend is (1.72$\pm$0.34)$\pi$ sr.\par
Note that if we replace the relativistic disc-line model with a narrow emission-line at 6.78 keV (in the source's rest-frame), as observed by \citet{brenneman09}, the broad excess between 6.1--6.7 keV (in the observer's frame) can not be modelled. Instead, we need at least three emission-lines to account for this excess resulting to an overparametrised physical model.  

\subsubsection{Ionized disc reflection with relativistic blurring (Model 4)}
\label{ssect:epic_spec_disc_reflect}
The simple relativistic disc-line component of Model 3 seems to give a good morphological description about the nature of the spectral excesses of Model 2. Nevertheless, the inclination angle, derived by the {\tt laor} model, seems to be unphysically high for a type $\rmn{I}$ Seyfert galaxy. Also, the centroid energy of the relativistic line strongly indicates that the inner disk is moderately ionized, yet no ionized Compton reflection component is included in Model 3. In this section we consider a relativistic self-consistent spectral-model, using as an input the reflection coming from a constant density illuminated disc. This model (Model 4) consists of the X-ray ionized-reflection model of \citet{ross05}, denoted as {\tt reflionx} in \textit{XSPEC}, convolved with the relativistic blurring multiplicative model of \citet{brenneman06}, denoted as {\tt kerrconv} in \textit{XSPEC}.\par
The ionized-reflection model takes into account several spectral signatures of ionized reflection in the source's rest-frame including: the Compton-hump at 30 keV, a strong ionized iron line at 6.7 keV, multiple emission-lines on top of a soft continuum between 0.3--3 keV, and a further extreme-\textit{UV} emission hump peaking at about 60 eV. The blurring model smears out the spectrum due to the strong gravity effects in a fully relativistic way and it allows the BH spin parameter, $\alpha$, to be a free fitting parameter.\par
During the fitting procedure, we keep all the parameters of Model 3 (i.e.\ the \fa and the \fb lines, as well as the {\tt pexrav} model-component) fixed at their best-fit values and we replace the disc-line component with the {\tt kerrconv$\times$reflionx}. We assume a disc of uniform emissivity extending from the radius of the innermost stable circular orbit, $R_{\rm ISCO}$, out to 400 $R_{\rm ISCO}$. The photon index of the ionized-reflection model-component, $\Gamma$, is tied to the photon index of the underlying redshifted-power-law model-component, representing the continuum.\par
\begin{figure}
\hspace*{0.61em}\includegraphics[width=3.428in]{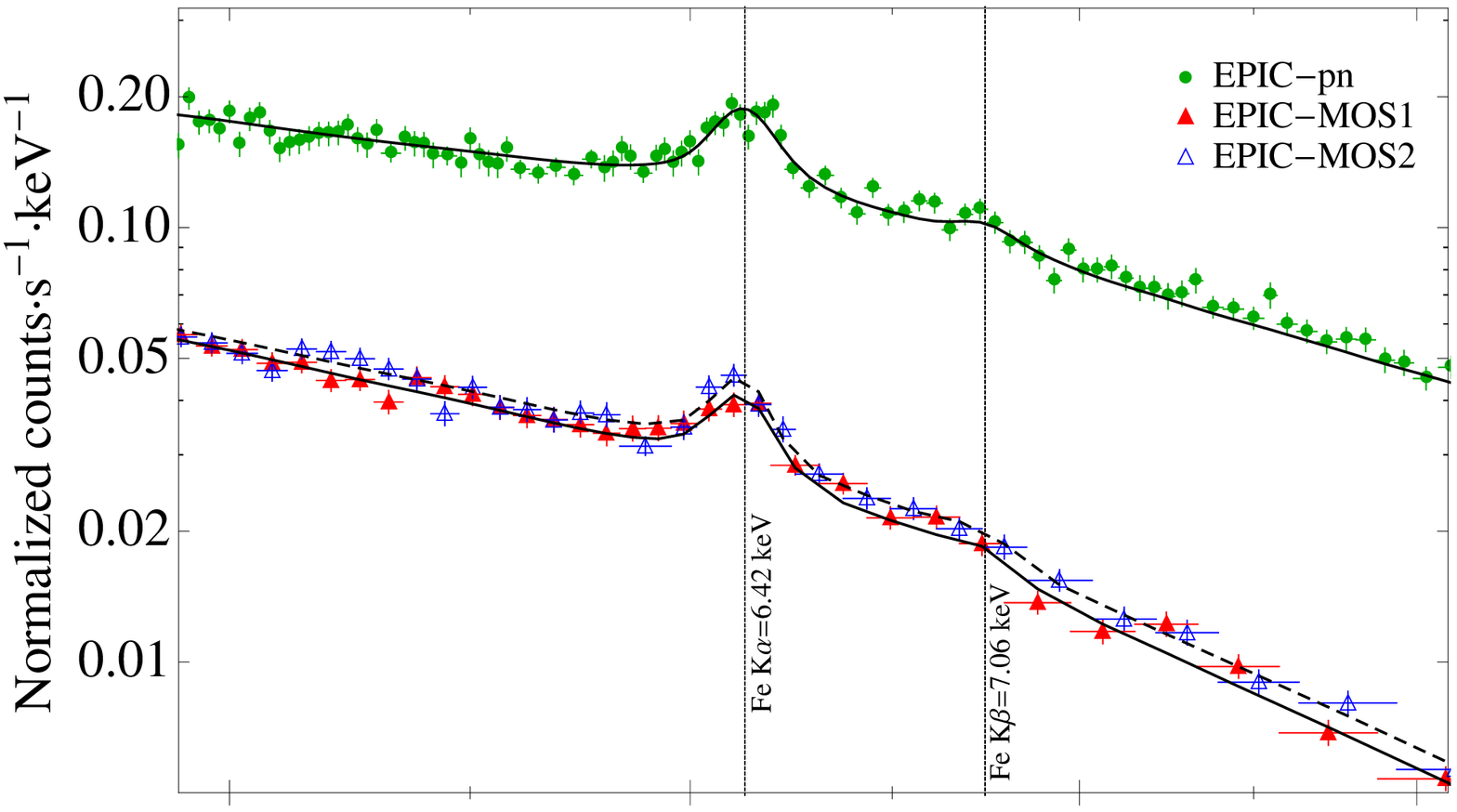}\\[-1.43em]
\hspace*{1.39em}\includegraphics[width=3.32in]{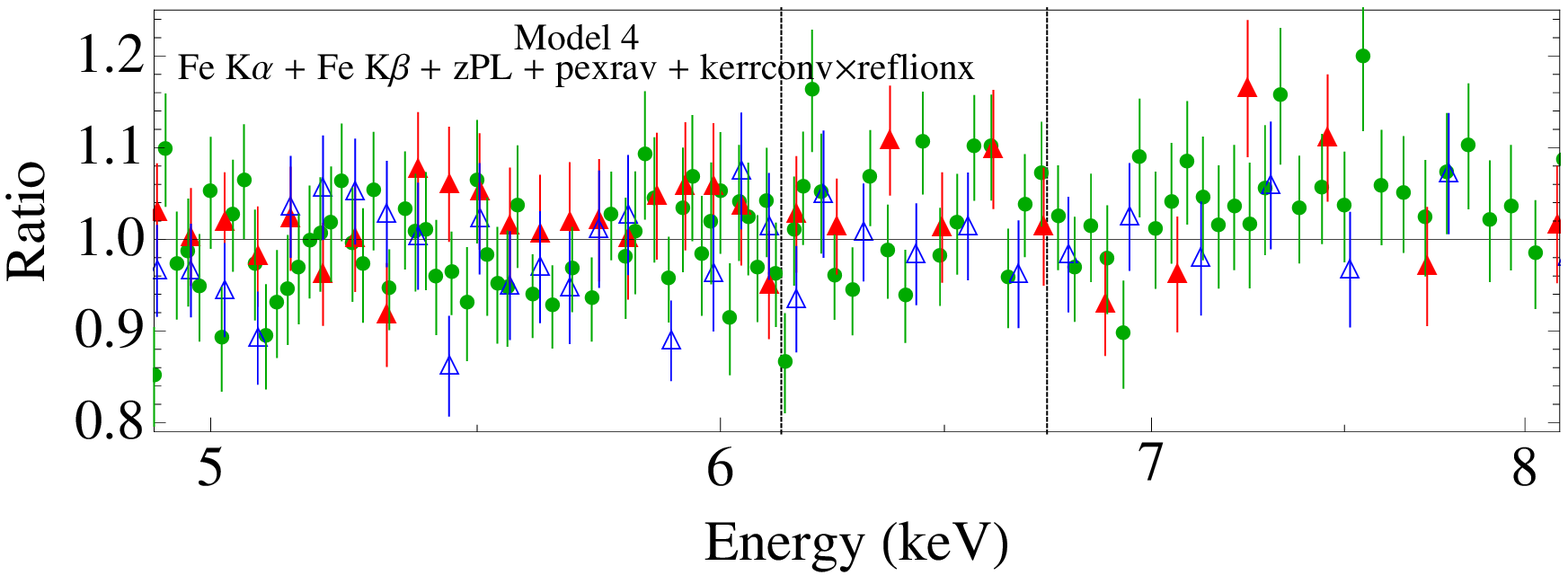}
\caption{Spectral fits in the 3--10 keV energy-range of Model 4. The vertical labels on the vertical lines indicate the energy in the source's rest-frame. Two narrow Fe lines --\fa (fixed), \fb (fixed)--, a reflection component (fixed), a redshifted power-law, and an ionized disc-reflection convolved with a relativistic blurring model.}
\label{fig:spectral4}
\end{figure}

The best-fit Model 4 gives a statistically acceptable fit for the X-ray EPIC-data of \f9 (Fig.~\ref{fig:spectral4}) yielding a $\chi^2$ value of 578.88 for 535 d.o.f. The best-fit model parameters are: power-law photon index of $\Gamma=1.84^{+0.02}_{-0.01}$, emissivity index of $q=2.07^{+0.12}_{-0.13}$, ionization parameter of $\xi=18.37^{+11.21}_{-6.32}$ erg cm s$^{-1}$, inclination of $\theta=50^{+18}_{-11}$ degrees, and BH spin-parameter of $\alpha=0.43^{+0.07}_{-0.06}$.\par Note that our data do not allow us to discern between uniform and non-uniform emissivity profiles. If we consider a {\tt kerrconv} model-component with a non-uniform emissivity profile, we get two emissivity indices of $q_1=2.65^{+2.68}_{-1.12}$ and $q_2=1.58^{+2.03}_{-0.96}$, and a break-radius of $r_{\rm br}=20^{+207}_{-15}$ $r_{\rm g}$. The two indices are consistent with the value derived previously for the case of uniform emissivity.

\subsection{The broad-band spectrum, 0.5--10 keV: Ionized disc-reflection with relativistic blurring (Model 4 (broad-band))}
\label{ssect:rel_blur_dis_refl_overll}
As we have mentioned above, at soft energies (i.e.\ below 2 keV) in addition to the power-law continuum, the ionized-reflection model predicts the presence of multiple emission-lines that are expected to be almost entirely blurred, due to the strong relativistic effects that operate in the vicinity of the central BH. In fact, it has been suggested that the so-called `soft-excess' emission in AGN could be the result of X-ray reflection from the inner parts of the accretion disc \citep[e.g.][]{crummy06}. For this reason, we examine whether Model 4 is able to reproduce the broad-band X-ray spectrum of the source in the 0.5--10 keV energy-band.\par
\begin{figure}
\hspace*{0.2em}\includegraphics[width=3.5in]{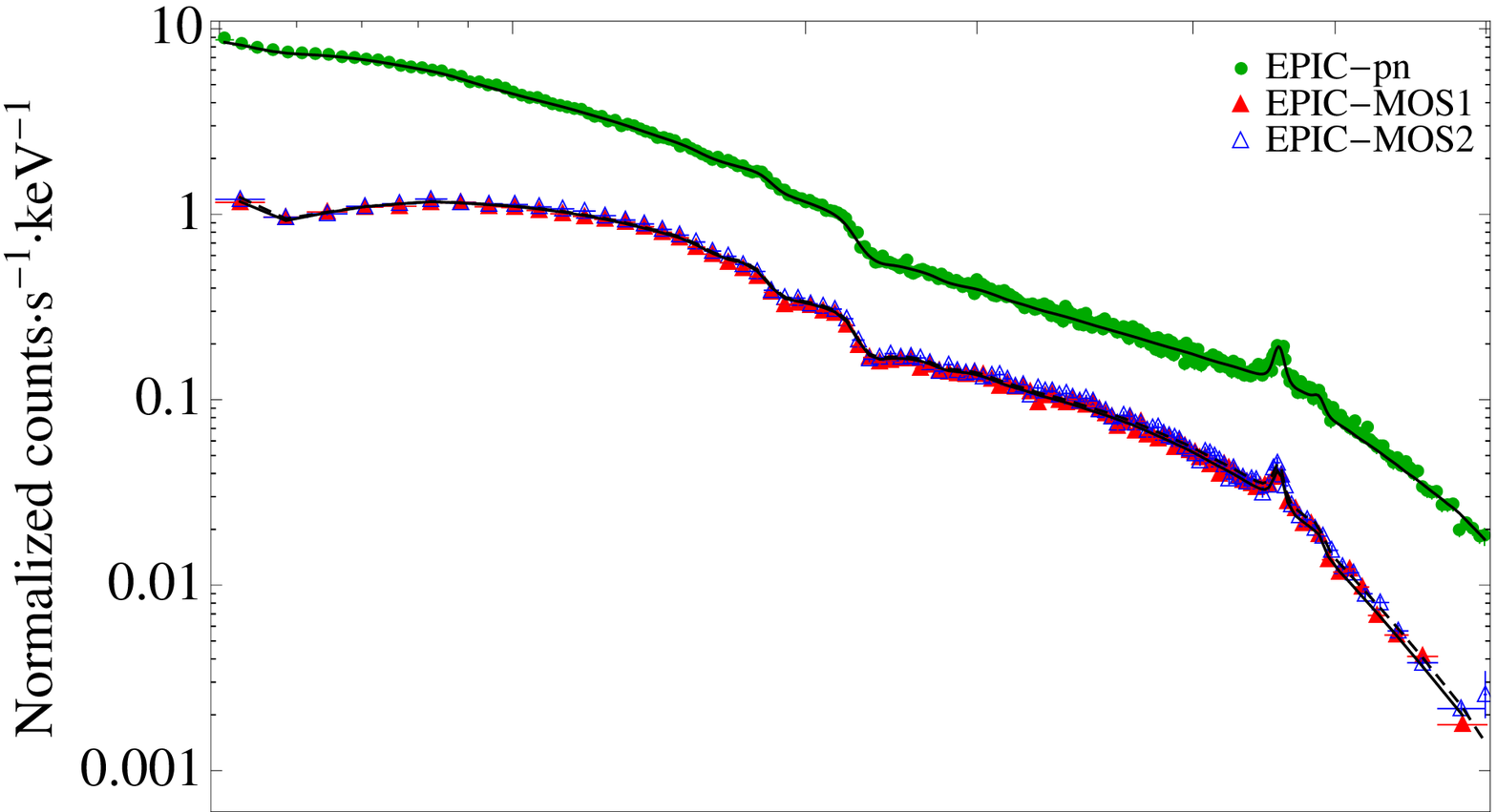}\\[-1.34em]
\hspace*{1.56em}\includegraphics[width=3.372in]{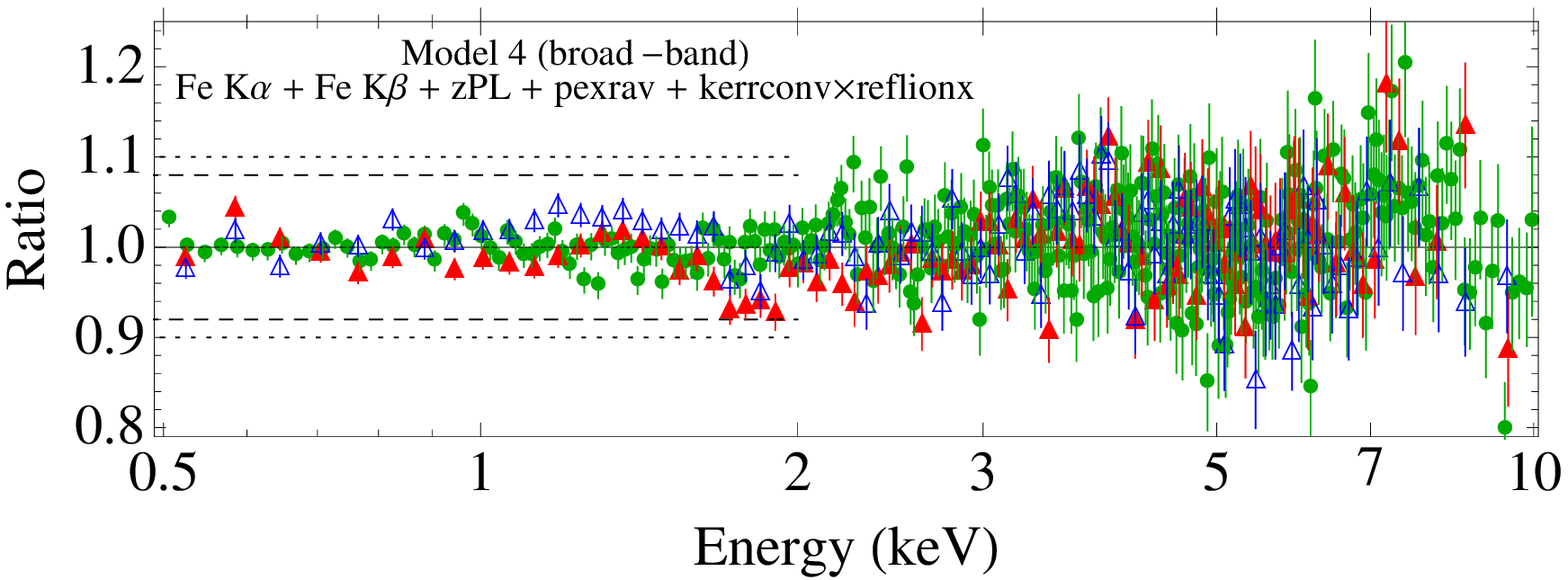}
\caption{Spectral fits in the 0.5--10 keV energy-range of Model 4. Two narrow Fe lines --\fa (fixed), \fb (fixed)--, a reflection component (fixed), a redshifted power-law, and an ionized disc-reflection convolved with relativistic blurring model. In the ratio plot, the area between the horizontal lines, in the 0.5--2 keV energy-range, indicate the 8  per cent (dashed-lines) and 10  per cent (dotted-lines) cross-calibration deviations between the pn and the MOS detectors.}
\label{fig:spectral5}
\end{figure}

To this end, we extend the energy-range down to 0.5 keV and we fit again Model 4 to the pn and MOS spectra simultaneously, having again the \fa and the \fb lines, as well as the {\tt pexrav} model-component fixed at their best-fit Model 3 values. The $\chi^2$ value of the fit is 1052.63 for 749 d.o.f. and the best-fit-model is shown in Fig.~\ref{fig:spectral5}. The best-fit model parameter values are: a power-law photon index of $\Gamma=2.01^{+0.01}_{-0.02}$, an emissivity index of $q=1.98^{+0.09}_{-0.05}$, an ionization parameter of $\xi=15.82^{+5.18}_{-3.26}$ erg cm s$^{-1}$, and an inclination angle of $\theta=64^{+7}_{-9}$ degrees. Finally, the best-fit BH spin parameter is $\alpha=0.39^{+0.48}_{-0.30}$. We outline the fact that the value of $\alpha$ is very sensitive to the rest of the fitting parameters. Even small deviations of the rest of the fitting parameters, around their best-fit values and within their quoted errors, can change significantly the result of $\alpha$.\par 
The value of the $\chi^2$ fit-statistic for this best-fit model is rather high. Strictly speaking, for the given d.o.f., the null hypothesis probability is extremely small, 1.3$\times10^{-12}$, implying that the particular best-fit model is highly unlikely to be a good representation of the underlying spectrum in the 0.5--10 keV energy-range. However, we believe that there are good reasons that this may not be the case due to the larger collecting area at lower energies. The quality of the fit is mainly determined by the data below 2 keV which have higher signal-to-noise ratio, than those above 2 keV, and thus bigger statistical-weight, during the estimation of $\chi^2$. At energies below 2 keV the cross-calibration uncertainties between the pn and the MOS detectors, as well as the individual calibration uncertainties of each detector separately are of the order of 8--10 per cent \citep[e.g.][]{mateos09}. The dashed- and the dotted-lines in the ratio plot of Fig.~\ref{fig:spectral5} indicate the 8 and 10 per cent residual-amplitude areas. We see that all of the ratio estimates, below 2 keV, are within these areas. What is even more interesting is that the pn and the MOS residuals do not typically agree with each other, as we would expect if indeed the model did not provide an acceptable representation of the observed spectra, suggesting that calibration inaccuracies dominate the $\chi^2$ fit-statistic. Finally, in order to limit the systematic calibration uncertainties to those originating only from the pn instrument, we ignore the MOS data and we fit the pn spectra alone. In this case, the $\chi^2$ of the overall fit is better (572.91 for 469 d.o.f.) but now due to the lower signal-to-noise ratio, since the MOS data have been excluded, we can not constrain adequately the fitting parameters e.g.\ the black-hole spin parameter is $\alpha=0.99^{+\mbox{--}}_{-0.98}$ and the inclination angle $\theta=90^{+ \mbox{--}}_{-81}$ degrees.\par
Furthermore, as we have shown in Sect.~\ref{ssect:hrs}, \f9 does show flux variations during this \textit{XMM-Newton} observation which are associated with spectral variations. In fact, the largest amplitude variations appear in the soft-band i.e.\ 0.5--1.5 keV. Therefore, it is not surprising that the fit-quality of (any) physically driven instantaneous spectral-model to the time-averaged spectrum may not be `acceptable'.\footnote{Obviously, it would be best to split the observation into smaller segments, e.g.\ three segments of 40 ks, and then fit the various models to the spectra of those segments. However, the source count-rate is not high enough in order to follow this approach and constrain the model parameters accurately.} For these reasons, judging from the best-fit residuals plot in Fig.~\ref{fig:spectral5}, we believe that Model 4 provides a broadly acceptable fit to the 0.5--10 keV spectrum of the source.\par 
The new best-fit broad-band photon index is steeper than the one obtained from the best-fit Model 4 in the 3--10 keV (Sect.~\ref{ssect:epic_spec_disc_reflect}). Interestingly, the new best-fit broad-band values for the remaining model parameters i.e.\ inclination angle, emissivity index, ionization parameter, and BH spin parameter, are entirely consistent (within the 90 per cent confidence limits) with those we have obtained from the best-fit Model 4 to the 3--10 keV spectrum.\par
\begin{figure}
\includegraphics[width=84mm]{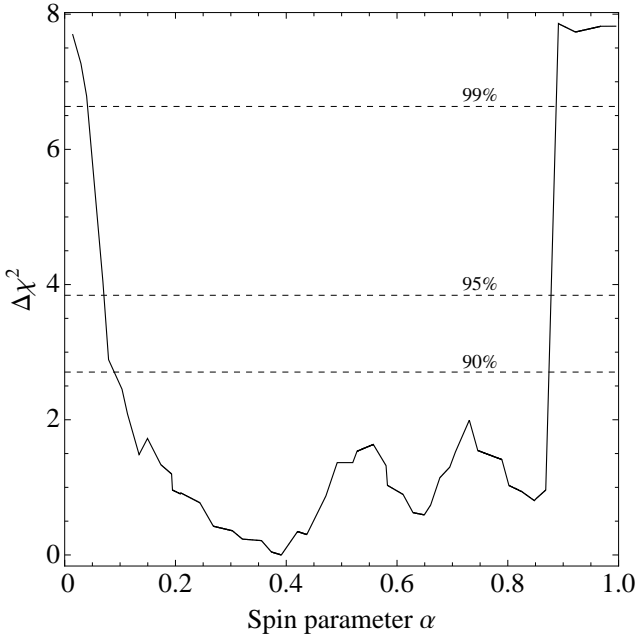}
\caption{$\Delta\chi^2$ plot versus the spin parameter $\alpha$ for the best-fit broad-band Model 4. The 90, 95 and 99 per cent confidence intervals for $\alpha$ are defined as the intersection points of the plot with the constant lines corresponding to $\Delta\chi^2$ of 2.706, 3.841 and 6.635, respectively.}
\label{fig:deltaChi}
\end{figure}

 
With respect to the BH spin-parameter, $\alpha$, Fig.~\ref{fig:deltaChi} shows the form of the $\Delta\chi^2$ plot versus $\alpha$, used in order to derive its uncertainties $\delta \alpha$, representing the 90 per cent confidence limit. Note here, a value of zero as well as the value of unity, are excluded at the 1 per cent significance level. We therefore conclude that relativistically blurred reflection from the innermost part of a slightly ionized disc, around a moderately spinning BH, may indeed be responsible for both the broad, asymmetric iron line and the `soft-excess' emission that we detect from \f9.\par
After estimating the best-fit parameters of Model 4 for the overall spectrum, we can compute the source X-ray flux and X-ray luminosity. In the 0.5--10 keV and 2--10 keV energy-ranges, the absorbed$[$unabsorbed$]$ flux is (2.5[2.6]$\pm$0.2)$\times10^{-11}$ erg cm$^{-2}$ s$^{-1}$ and (1.5[1.5]$\pm$0.1)$\times10^{-11}$ erg cm$^{-2}$ s$^{-1}$, respectively. The corresponding X-ray luminosities for the same energy-bands are (1.26[1.33]$\pm$0.03)$\times10^{44}$ erg s$^{-1}$ and (0.76[0.77]$\pm$0.05)$\times10^{44}$ erg s$^{-1}$, respectively (the flux and luminosity errors correspond to one standard deviation in order to be easily comparable with previous observations). Finally, after applying the bolometric luminosity correction of \citet{marconi04} to the derived 2--10 keV X-ray luminosity, we get for \f9 a bolometric luminosity of 2.34$\times 10^{45}$ erg s$^{-1}$. This value corresponds to an Eddington luminosity ratio of 0.07, comparable to the one derived by \citet{woo00} (see Sect.~\ref{sect:intro}) but smaller by a factor of two than that derived by \citet{schmoll09}.

\begin{table*}
\centering
\begin{minipage}{140mm}
\caption{Best-fit and fixed parameters for the X-ray spectral Models 2, 3 and 4 in the 3--10 keV energy-range and for Model 4 (broad-band) in the 0.5--10 keV energy-range.}
\label{tab:spec_modelFits}
\begin{tabular}{@{}llcccc}
\hline
Number & Model-parameters & Model 2 & Model 3 & Model 4 & Model 4 (broad-band) \\
\hline
 & Redshift, z & 0.047 & 0.047 & 0.047 & 0.047 \\
1 & {\tt wabs}: Column density, $N_{\rm H}$ (cm$^{-2}$) & 3.20$\times10^{20}$ &  3.20$\times10^{20}$ & 3.20$\times10^{20}$ & 3.20$\times10^{20}$\\
2 & {\tt zGauss}: \fa line-energy (keV) & 6.42$\pm$0.01& 6.42 & 6.42 & 6.42\\
3 & {\tt zGauss}: \fb line-energy (keV) & 7.06  & 7.06 & 7.06 & 7.06\\
4 & {\tt zpowerlw}: Photon index, $\Gamma$ & 1.85$^{+0.03}_{-0.02}$ & 1.97$^{+0.07}_{-0.06}$ & 1.84$^{+0.02}_{-0.01}$ & 2.01$^{+0.01}_{-0.02}$\\
5a & {\tt pexrav}: Photon index, $\Gamma_{\rm pex}$ & 1.87$\pm$0.03 & 1.87 & 1.87 & 1.87\\
5b & {\tt pexrav}: Relative reflection, $R_{\rm pex}$ & 0.87$^{+0.05}_{-0.06}$ & 0.87 & 0.87 & 0.87\\
5c & {\tt pexrav}: Energy cut-off, $E_c$ (keV) & 200 & 200 & 200 & 200\\
5d & {\tt pexrav}: Elem. Z$>$2 abundance & 1 & 1 & 1 & 1\\
5e & {\tt pexrav}: Iron abundance & 0.8 & 0.8 & 0.8 & 0.8\\
5f & {\tt pexrav}: Inclination, $\phi$ (degrees) & 40 & 40 & 40 & 40\\
6a & {\tt laor}: Line-energy (keV) &  & 6.75$\pm$0.18 & & \\
6b & {\tt laor}: Emissivity index, $q$ &  & 2.46$^{+0.40}_{-0.37}$ & & \\
6c & {\tt laor}: Inclination, $\theta$ (degrees) &  & 84$^{+2}_{-17}$ &  &  \\
6d & {\tt laor}: Inner radius ($r_{\rm g}$) &  & 6 &  & \\
6e & {\tt laor}: Outer radius ($r_{\rm g}$) &  & 300 &  & \\
7a & {\tt kerrconv}: Emissivity index, $q$ &  &  & 2.07$^{+0.12}_{-0.13}$ & 1.98$^{+0.09}_{-0.05}$\\
7b & {\tt kerrconv}: Inclination, $\theta$ (degrees) &  &  & 50$^{+18}_{-11}$ & 64$^{+7}_{-9}$\\
7c & {\tt kerrconv}: Spin parameter, $\alpha$ &  &  & 0.43$^{+0.07}_{-0.06}$ & 0.39$^{+0.48}_{-0.30}$\\
7d & {\tt kerrconv}: Inner radius, ($r_{\rm ISCO}$) &  &  & 1 & 1\\
7e & {\tt kerrconv}: Outer radius, ($r_{\rm ISCO}$) &  &  & 400 & 400\\
8a & {\tt reflionx}: Iron abundance &  &  & 0.8 & 0.8\\
8b & {\tt reflionx}: Photon index, $\Gamma$ &  &  & Tied to param. 4 & Tied to param. 4 \\
\multirow{2}{*}{8c} & {\tt reflionx}: Ionization parameter, $\xi$  & \multirow{2}{*}{ } & \multirow{2}{*}{ } & \multirow{2}{*}{18.37$^{+11.21}_{-6.32}$} & \multirow{2}{*}{15.82$^{+5.18}_{-3.26}$} \\
                    &\hspace{4.6em}(erg cm$^{-2}$ s$^{-1}$) &   &  &  & \\
\hline
$\chi^2/$d.o.f. &  & 600.29$/$538 & 559.18$/$536 & 578.88$/$535 & 1052.63$/$749 \\
\hline
\end{tabular}
\medskip
Values without uncertainties correspond to fixed fitting parameters. 
\end{minipage}
\end{table*}

\section{Summary and Discussion}
\label{sect:discus}
We have analysed a long observation of \f9, obtained with the \textit{XMM-Newton} observatory during December 2009. EPIC light-curves have been produced and studied in several energy-bands, a thorough X-ray spectral analysis has been performed on both the RGS and the EPIC data-sets.\par
The hardness-ratio analysis has revealed a `softer-when-brighter' source-behaviour which is typical for radio-quiet type $\rmn{I}$ Seyfert galaxies. The observed spectral variations do not have a large amplitude, In particular, the 5--10 keV light-curve (Fig.~\ref{fig:lightcurves}, lower panel) shows a small amplitude flux-increase of the order of 10 per cent and a small fractional variability amplitude, 2.2$\pm$2.2 per cent. This result implies that our best-fit model results for the 3--10 keV spectral region should not be strongly affected by significant spectral variations which could make the study of the average spectrum problematic.\par
However, this is not the case for the lower energy-part of the spectrum between 0.5--2 keV (i.e.\ the `soft-excess' spectral region) where the 0.5--1.5 keV light-curve (Fig.~\ref{fig:lightcurves}, top panel) exhibits an overall flux increase of 30 per cent and a fractional variability amplitude of 7.3$\pm$0.1 per cent. This indicates that any model fit to the overall broad-band average-spectrum will be affected by these variations. The 0.5--2 keV data have smaller uncertainties with respect to those in the 2--10 keV, and thus the soft-band data drive the overall fit by having larger statistical weight. In fact, these differences in the amplitude variations between the soft and the hard X-ray bands can possibly account for the statistically low-quality fit of Model 4 to the overall spectrum of the source.\par
The main results from our X-ray spectral analysis can be summarized as follows:
\begin{enumerate}
\item Our view to the central source in \f9 does not appear to be affected by significant absorption either from neutral or ionized material.
\item Apart from a strong narrow iron emission-line and an absorption edge, we have also detected a broad emission feature in the 6.1--7.7 keV energy-band, which can be well explained with relativistically blurred reflection from the innermost part of a moderately ionized disc, around a moderately spinning BH.
\end{enumerate}\par
Below, we discuss some implications of these two main results, within the context of current theoretical models of the X-ray emission in AGN.

\subsection{On the absence of intrinsic absorption in \f9}
\label{ssect:absence_absorption}
Early observations by EXOSAT and GINGA showed that around 50 per cent of all type $\rmn{I}$ Seyfert galaxies show strong evidence of warm-absorption \citep{turner89,nandra94}. Moreover ASCA observations suggested an even higher fraction of around 50--70 per-cent \citep{reynolds97,george98}. Finally, \citet{blustin05}, based on an \textit{XMM-Newton} sample of 23 type $\rmn{I}$ Seyfert, found 17 objects having intrinsic ionized absorption, probably originating in outflows from the dusty torus.\par
Nevertheless, there is a significant population of type $\rmn{I}$ Seyfert galaxies that does not show any indication of strong X-ray warm-absorption features like: Akn\,120 \citep{vaughan04}, Akn\,564 \citep{papadakis07}, Mrk\,205 \citep{reeves01}, Mrk\,335 \citep{gondoin02}, Mrk\,478 \citep{marshall03}, MCG--2-58-22 \citep{weaver95}, PKS 0558-504 \citep{papadakis10}, and TON\,S\hspace*{-0.2em} 180 \citep{turner01,vaughan02} (but for this source see also \citet{rozanska04} for very weak absorption features).\par
Our main result, from the spectral analysis of the RGS data, is that there is no indication of significant X-ray warm-absorbing material along the line of sight to the central nucleus of the \f9. This is in agreement with the results of \citet{gondoin01}. Even the 99.73 upper confidence limit of N$_{\rm H}$, 3.72$\times 10^{20}$ cm$^{-2}$ (Fig.~\ref{fig:wa_contour}), is smaller then the column density of the warm-absorbing material in most type $\rmn{I}$ Seyfert galaxies \citep[e.g.][]{blustin05}. This implies that our observations are not `contaminated' by additional material between the central engine and us, making \f9 an ideal source for studying the physical properties of the immediate environment of the BH.\par 
The reason for the lack of warm-absorbing material in \f9 (and the other sources listed above) remains still not fully understood.  One possibility is that a significant column density of ionized gas exists but it is too highly ionized to show significant spectral features. We marginally detected weak emission-lines from ionized oxygen for which $\log_{\rm 10}\left(U\right)=-1.1$. This is a rather small value, and certainly not indicative of the presence of highly ionized material in the vicinity of Fairall\,9's nuclear region. Another possibility is that the covering fraction of the absorbing material is less than unity. This could naturally arise in the case when the absorbing material is associated with a disc-wind which rises perpendicularly from the disc at a narrow range of radii \citep[e.g.][]{elvis00}.\par
Both the significant outflow velocity of the emitting gas, as well its main source of ionization, as inferred by the O\,{\sc vii}\,K$\alpha$ (f/r) ratio, 4.3$^{+4.1}_{-2.4}$ (i.e.\ photo-ionization), suggest that the emitting gas in \f9 could be probably part of a nuclear wind driven out by radiation pressure. This same evidence is found in numerous other Seyfert galaxies, both absorbed \citep{guainazzi08} and unabsorbed \citep{bianchi09}. It is commonly interpreted as a bi-conical outflow, launched from the innermost region of the AGN and spatially co-located with the optical ionization cones frequently seen in  O\,{\sc iii} emission in type $\rmn{II}$ Seyfert galaxies \citep{ogle00,bianchi06}.

\subsection{The `bare' central nucleus of \f9}
\label{ssect:clear_view}
Our X-ray spectral results from the analysis of the pn and MOS data of \f9, are in full agreement with the `canonical' picture for the central region in Type I AGN, as it has been established the last 20 years. For example, the best-fit power-law continuum slope-estimate, coming from fitting Model 4 to the broad-band data in the 0.5--10 keV energy-range, is $\Gamma=2.01^{+0.01}_{-0.02}$. This value is slightly softer than the one derived by \citet{gondoin01}, 1.80$^{+0.25}_{-0.13}$ using earlier \textit{XMM-Newton} data, (but still in accordance within the errors), and harder from the one reported by \citet{schmoll09}, 2.09$\pm0.01$ using \textit{Suzaku} data. This is absolutely in accordance with the spectral variability behaviour of the source as a function of flux. During this \textit{XMM-Newton} observations the source flux in the 0.5--10 keV energy-range is around 2.5$\times$10$^{-11}$ erg cm$^{-2}$ s$^{-1}$ (Sect.~\ref{ssect:rel_blur_dis_refl_overll}). \citet{gondoin01} derived an equal value for the flux, in contrast to \citet{schmoll09} who derived a higher flux of 4$\times10^{-11}$ erg cm$^{-2}$ s$^{-1}$. Therefore, given the `softer-when-brighter' source-behaviour, it is not surprising that the \textit{XMM-Newton} spectral slope is slightly harder than that derived from the \textit{Suzaku} data.\par
Moreover, we have detected a narrow emission iron line at 6.42 keV (in the source's rest-frame). The best-fit value of the eqw for this line, is 103$\pm$19 eV which is somehow lower from the ones derived by \citet{gondoin01} and \citet{schmoll09}, 120 eV (no errors are quoted) and 130$\pm$10 eV, respectively, but still within the quoted errors. Nowadays, narrow iron lines, at 6.4 keV (in the source's rest-frame), are detected almost universally in bright AGN X-ray spectra. The eqw of this line in our spectrum is very consistent to those observed in other sources. For example, \cite{bianchi09}, studying the spectra of 81 AGN, estimated an average eqw of the narrow \fa line of 76$\pm$6 keV. Similar results have also been reported by \citet{nandra07} and \citet{calle10}.\par
This narrow iron line is usually attributed to reprocessing from cold Compton-thick matter which is most probably associated with the putative, distant molecular-torus in Seyfert galaxies \citep[e.g.][]{matt91}. Based on our best-fit results of Model 3, we find a reflection fraction of $R=0.86\pm 0.17$. From our estimate, the distant reflector should subtend an apparent solid angle of around $1.7\pi$ sr at the X-ray source. However, a more accurate statement about this and about the opening angle of the torus, can be made only by fitting to the data a physical model of the torus itself, rather than a slab (as we have done by using the \textit{XSPEC} model {\tt pexrav}) to account for the continuum reflection associated with the narrow iron line.\par
Following the best-fit linear model \citep{bianchi07} of the Iwasawa-Taniguchi (IT) effect \citep{iwasawa93}, indicating the scenario of anticorrelation between the narrow \fa emission-line and the 2--10 keV luminosity, we expect an eqw of 56$\pm4$ eV (for a source luminosity in 2--10 keV energy-range of 0.77$\times 10^{44}$ erg s$^{-1}$, Sect.~\ref{ssect:rel_blur_dis_refl_overll}). In order to check the consistency between our direct estimates and the value predicted by the IT effect, we perform a $z$-test. The value of ${\itl z}$-statistic is -4.38\footnote{Note that the standard deviation of the estimated eqw of \fa line is 10 eV}, having a chance coincidence probability of only 1.17$\times10^{-5}$. That means that with more than 99.99 per cent confidence we can reject the null hypothesis that the measured eqw of the \fa line is comparable to that predicted by the IT relation.\par
The broad residuals of Model 2 around 6.1--7.7 keV (Fig.~\ref{fig:spectral2_3}, top panel), can be modeled very well by a relativistically blurred disc-reflection model. A similar broad-emission spectral feature has also been observed by \textit{Suzaku} from the same source \citep{schmoll09}. Due to the lack of any significant warm-absorption in the nucleus of \f9, this picture should unveil directly the physical properties prevailing in the immediate environment of the BH. What is even more interesting is the fact that the same relativistically blurred disc-reflection model can also account satisfactorily, under certain assumptions (see Sect.~\ref{ssect:rel_blur_dis_refl_overll}), for the `soft-excess' emission that we also detect from \f9.\par
The emissivity index that we have derived from the best-fit Model 4 to the overall X-ray spectrum is $q=1.98^{+0.09}_{-0.05}$. \citet{schmoll09} found an upper bound of 5 but more stringent constraints have been derived by \citet{patrick11}, $2.7^{+0.7}_{-0.4}$, which agrees reasonably well with our findings.\par
Our best-fit value for the BH spin parameter for \f9, $\alpha=0.39^{+0.48}_{-0.30}$, is in accordance with the one derived by \citet{schmoll09}, $\alpha=0.60\pm0.07$, using the \textit{Suzaku} data. The latter value comes from a broader-band spectral-fit in the 0.7--10 keV (X-ray Imaging Spectrometer) and 12--30 keV (Hard X-ray detector/PIN) whereas ours has been derived from a smaller spectral-range of 0.5--10 keV. Note that after ignoring the spectral measurements below 2 keV, \citet{schmoll09} got a value of $\alpha=0.5^{+0.1}_{-0.3}$. Also, \citet{patrick11}, used the same \textit{Suzaku} data of \f9, together with the hard X-ray data from the BAT, on board the \textit{SWIFT} satellite. By performing spectral-fits in the 0.5--100 keV energy-range (excluding 2.7--1.95 keV and 10--15 keV) they obtained a value of $\alpha=0.44^{+0.04}_{-0.11}$ which is even more consistent with our estimate.\par
Recent theories studying the effects of chaotic accretion episodes on the BH spin-parameter \citep{king08}, predict that $\alpha$ should peak at around 0.1--0.2 for a BH mass of $10^8\;{\rm M}_{\sun}$, depending on the vertical viscosity of the accretion disc. On the other hand, prolonged accretion should result in much larger spin-parameters, of the order of $\alpha>0.9$
\citep{berti08}. Consequently, the BH spin-parameter of around 0.4, that we found for \f9, is more consistent with models which favour a chaotic accretion scenario.\par  
Up to now, rather few studies have been conducted to specify the BH spin-parameter, $\alpha$, for the type $\rmn{I}$ Seyfert galaxies. A value of $\alpha=0.989^{+0.009}_{-0.002}$ has been reported for MCG--6-30-15 \citep{brenneman06} and \citep{miniutti09} found for SWIFT\,J2127.4+5654 a value of $\alpha=0.6\pm0.2$. Moreover \citet{zoghbi10} found a high BH spin parameter value of $\alpha>0.98$ for 1H\,0707-495 and \citet{gallo10} derived a value of 0.7$\pm0.1$ for Mrk\,79. Finally, \citet{patrick11} suggested intermediate BH spin-parameters for Mrk\,335 and NGC\,7469 of 0.70$^{+0.12}_{-0.01}$ and 0.69$\pm0.09$, respectively.\par
This long-look \textit{XMM-Newton} observation of \f9 has revealed that we are dealing with a `clean' source since our X-ray spectrum is not affected by significant warm-absorption. Having this uncontaminated view of the central engine, a single relativistic disc-reflection model can explain the overall emission from 0.5 keV up to 10 keV. The existence of a moderately spinning Kerr BH is strongly favoured from our data consolidating firmly that this source is in the intermediate spin-class of AGN.

\section*{Acknowledgments}
DE and IMM acknowledge the Science and Technology Facilities Council (STFC) for support under grant ST/G003084/1. IP and FN acknowledge support by the EU
FP7-REGPOT 206469 grant. PA acknowledges support from Fondecyt grant-number 1110049. This research has made use of NASA's Astrophysics Data System Bibliographic Services. Finally, we are grateful to the anonymous referee not only for the very prompt review, but also for the useful comments and suggestions that helped improved the quality of the manuscript substantially.

\bibliographystyle{mnauthors}

\bsp
\label{lastpage}
\end{document}